\documentclass[
superscriptaddress,
amsmath,amssymb,
aps, twocolumn, prb, floatfix
]{revtex4-2}

\usepackage[utf8]{inputenc}
\usepackage[T1]{fontenc}
\usepackage{mathptmx}
\usepackage{tabularx}
\usepackage{booktabs} 
\usepackage{array}    
\usepackage[italian,english]{babel}
\usepackage[dvipsnames]{xcolor}

\usepackage[breaklinks,pdftex,pdfpagelabels,bookmarks]{hyperref}
\hypersetup{colorlinks=true,linktocpage=true,pdfstartpage=3,pdfstartview=FitH,breaklinks=true,pdfpagemode=UseNone,pageanchor=true,pdfpagemode=UseOutlines,plainpages=false,bookmarksnumbered, bookmarksopen=true,bookmarksopenlevel=1,hypertexnames=true,urlcolor=Brown,linkcolor=MidnightBlue!60!Black,citecolor=PineGreen}
\usepackage{cleveref}
\usepackage[version=4]{mhchem}
\usepackage{physics}
\usepackage{siunitx}
\usepackage{comment}
\usepackage{graphicx}
\usepackage{dcolumn}
\usepackage{bm}
\usepackage{amsmath}
\usepackage{mathtools, nccmath}
\usepackage{amsfonts}
\usepackage{amssymb}
\usepackage{xcolor}
\usepackage{bbold}

\usepackage{multirow}
\graphicspath{{./Images/}}

\begin{document}
\preprint{APS/123-QED}
\title{$0-\pi$ transitions in non-Hermitian magnetic Josephson junctions}

\author{R. Capecelatro}
\email{rcapecelatro@unisa.it} 
\affiliation{Dipartimento di Fisica "E.R. Caianiello"$,$ Università di Salerno$,$ via Giovanni Paolo II$,$ 132$,$ I-84084$,$ Fisciano (SA)$,$ Italy}
\affiliation{INFN$,$ Sezione di Napoli$,$ Gruppo collegato di Salerno$,$ Italy}
\author{M. Marciani}

\affiliation{Dipartimento di Fisica E. Pancini$,$ Università degli Studi di Napoli Federico II$,$ Monte S. Angelo$,$ via Cinthia$,$ I-80126 Napoli$,$ Italy}
\author{C. Guarcello}
\affiliation{Dipartimento di Fisica "E.R. Caianiello"$,$ Università di Salerno$,$ via Giovanni Paolo II$,$ 132$,$ I-84084$,$ Fisciano (SA)$,$ Italy}
\affiliation{INFN$,$ Sezione di Napoli$,$ Gruppo collegato di Salerno$,$ Italy}
\author{G. Campagnano}
\affiliation{CNR-SPIN$,$ UOS Napoli$,$ Monte S. Angelo$,$ via Cinthia$,$ I-80126 Napoli$,$ Italy}
\author{P. Lucignano}
\affiliation{Dipartimento di Fisica E. Pancini$,$ Università degli Studi di Napoli Federico II$,$ Monte S. Angelo$,$ via Cinthia$,$ I-80126 Napoli$,$ Italy}
\author{R. Citro}
\affiliation{Dipartimento di Fisica "E.R. Caianiello"$,$ Università di Salerno$,$ via Giovanni Paolo II$,$ 132$,$ I-84084$,$ Fisciano (SA)$,$ Italy}
\affiliation{INFN$,$ Sezione di Napoli$,$ Gruppo collegato di Salerno$,$ Italy}
\affiliation{CNR-SPIN$,$ c/o Università di Salerno$,$ I-84084 Fisciano (Salerno)$,$ Italy}

\begin{abstract}
We study the transport properties of non-Hermitian magnetic Josephson junctions, considering a superconductor–quantum dot–superconductor device coupled to a ferromagnetic metallic reservoir in the presence of an external magnetic field. We focus on the $0-\pi$ transitions that occur when the equilibrium phase difference between the superconductors shifts from $\phi=0$ to $\phi=\pi$ upon increasing the magnetic field amplitude. The coupling to the environment induces spin-dependent dissipation and leads to the broadening of the junction Andreev levels. 
By combining Green’s function calculations with an effective non-Hermitian description restricted to the sub-gap Andreev quasi-bound states, we show that dissipation shifts the $0-\pi$ transition to higher magnetic fields. 
Remarkably, also the relative angle between the applied field and the reservoir magnetization can be used to drive the transition, at fixed field magnitude. We demonstrate that this effect can be entirely ascribed to the behavior of the complex eigenvalues of the non-Hermitian Hamiltonian. 
These findings highlight non-Hermiticity as a resource that can introduce new control knobs for engineering the current–phase relation in superconducting junctions.
\end{abstract}
\maketitle

\section{\label{sec: intro}Introduction}

Non-Hermitian (NH) Hamiltonians provide an effective description of open quantum systems~\cite{BreuerPetruccione, Weiss}, where losses and gains are incorporated at the level of complex eigenenergies~\cite{Gong_2020, Bender07, Bender:1998, Kawabata2019}. The emergence of exceptional points (EPs) in the spectrum~\cite{Zhen2015, Shen2018, Cerjan2019, Bergholtz2021, Oku23} and the NH skin effect~\cite{FoaTorres2018, Yao18, Zhang2020, Zhang:2022, ZhangKai2022} has been widely explored in optical systems~\cite{Berry2004, Pen14, Doppler2016, St-Jean2017, Zhou2018, El-Ganainy2019, schonleber2016, Xu2016}. More recently, by leveraging on different loss channels, NH phenomena have also been investigated in condensed matter platforms~\cite{Mandal2020, Zhang:2021, Geng2023, Qing2024, Philip2018, Chen2018, Bergholtz2019, Budich2020, Cayao2023_2, Mi14, SanJose2016, Avila:2019, Cayao2023, Arouca2023, Sayyad2023, Cayao2024, CayaoAguado2024, Kawabata2018, Cayao2022, Kornich2022, Kornich2022_2, Kornich2023, Kokhanchik2023, Paya2026}.

In mesoscopic systems, dissipation arising from the coupling to external environments is often unavoidable and can strongly affect both spectral and transport properties. Of particular interest are the effects of fermionic baths on Josephson junctions (JJs), which represent the building blocks of superconducting quantum technologies~\cite{Josephson1962, Barone1982, Devoret1984, Martinis1987, Clarke1988, Cleland1988, Devoret2013}.
In this context, normal metal leads in multi-terminal junctions provide an effective way to model fermionic reservoirs, with the junction–bath interaction described by an imaginary and frequency-independent term in the JJ effective NH Hamiltonian~\cite{Li2024, Shen2024, Beenakker2024, CayaoSato2024, CayaoSato2024_2, Capecelatro2025, Pino2025, Ohnmacht2025, Solow2025, Li2025, Junjie2025, Capecelatro2026, CayaoSato2026}.
The coupling to the environment leads to a phase-dependent broadening of the JJ Andreev levels, whose eigenstates are thus addressed as Andreev \emph{quasi-bound} states (quasi-ABS).
The levels broadening results in an attenuation of the Andreev current, proportional to the phase-derivative of the levels energy, and in additional current contributions originating entirely from the phase-derivative of the levels width~\cite{Beenakker2024, Shen2024, Capecelatro2025, Pino2025, Capecelatro2026}.

While the equilibrium transport properties of NH JJs, including the role of EPs in the Andreev spectrum, have been extensively studied, much less is known about the controlled engineering of specific current–phase relations (CPR) in open junctions. This includes $0-\pi$ transitions~\cite{Sellier2005, Benjamin2007, Asano2019,  Minutillo2021, Ahmad2022, Capecelatro2023}, the anomalous Josephson effect~\cite{Reynoso2008, Zazunov2009, Yokoyama2013, Yokoyama2014, Campagnano_2015, Minutillo2018, Guarcello2020, Strambini2020}, and the Josephson diode effect~\cite{Buzdin2008, Ando2020, Kou2021, Akito2022, Davydova2022, Trahms2023, Nadeem2023, Guarcello2024, Maiellaro2024}. While a NH Josephson diode has been proposed within a SQUID geometry including a NH quantum dot JJ~\cite{Junjie2025}, the impact of fermionic baths on junctions exhibiting such CPR behaviors has not yet been systematically addressed. Even more interestingly, non-Hermiticity naturally introduces new degrees of freedom to control the supercurrent, possibly providing a different resource to tailor the junction CPR.

In this respect, we investigate the phenomenology of $0-\pi$ transitions in NH magnetic JJs. In conventional junctions with magnetic barriers, tuning the magnetic field amplitude or the length of the weak link can induce a sign reversal of the supercurrent, shifting the equilibrium phase difference from $\phi=0$ to $\phi=\pi$~\cite{Buzdin1982, Ryazanov2001, Golubov2004, Buzdin2005, Bergeret2005}. Such $\pi$ junctions are appealing for applications in SQUID-based architectures for self-biased flux qubits ($\pi$-qubits), with potential advantages from a scalability perspective~\cite{Ioffe1999, Blatter2001, Majer2002, Ustinov2003, Yamashita2005, Yamashita2006, Yamashita2020}.

Moreover, in magnetic junctions with spin–orbit interactions, $0-\pi$ transitions are closely related to the emergence of large anomalous Josephson currents and Josephson diode effects~\cite{Reynoso2008, Zazunov2009, Yokoyama2013, Yokoyama2014, Campagnano_2015, Minutillo2018} being key ingredients for CPR engineering in such devices.

Here, we study the $0-\pi$ transition in a minimal setup consisting of a superconductor–quantum dot–superconductor junction coupled to a ferromagnetic metallic reservoir, which induces spin-dependent level broadening. Combining Green’s function (GF) techniques and the NH current formula recently derived in Refs.~\cite{Capecelatro2025, Capecelatro2026}, we analyze the Zeeman field driven $0-\pi$ transitions and we report two main results.

First, dissipation shifts the transition to higher magnetic fields by a counterintuitive increase of current at certain phase-biases, which makes the $0$ state more robust. Second, the transition can be tuned through the relative angle between the applied magnetic field and the reservoir magnetization, providing a simple control knob for the JJ phase. 

We show that the phenomenology of the $0-\pi$ transitions in such devices can be understood in terms of the junction quasi-ABS. Although the supra-gap continuum should be included within a GF framework to compute the exact Josephson current, the NH Hamiltonian formalism already provides a comprehensive picture of the $0-\pi$ transition and its tunability.
Our findings testify that controlled $0-\pi$ Josephson devices can be engineered also with open JJs. The possibility of reaching stable $\pi$ states in such devices may thus be promising for the realization of $\pi$-qubits~\cite{Ioffe1999, Blatter2001, Majer2002, Ustinov2003, Yamashita2005, Yamashita2006, Yamashita2020} also in moderately noisy superconducting circuits. Moreover, our results indicate that losses, rather than being only detrimental, can increase the current and unlock different control parameters for CPR engineering in NH superconducting devices.

This work is organized as follows. In Sec.~\ref{sec: Model} we recall how to derive the NH Hamiltonian of the system and how to compute the Andreev levels current. We analyze in Sec.~\ref{sec: 0_pi_N_lead} the impact of a normal metal (N) reservoir on the $0-\pi$ transition. In Sec.~\ref{sec: 0_pi_F_lead} we consider the case of spin-dependent dissipation, i.e. a ferromagnetic reservoir (F), and investigate the tunability of the $0-\pi$ transitions with the relative angle between the magnetic field and the bath magnetization. Finally, in Sec.~\ref{sec: conclusions} we summarize our results and discuss possible future perspectives. 
  \section{Model}
    \label{sec: Model}
    \begin{figure}[b]
		\centering
		\includegraphics[scale=0.5]{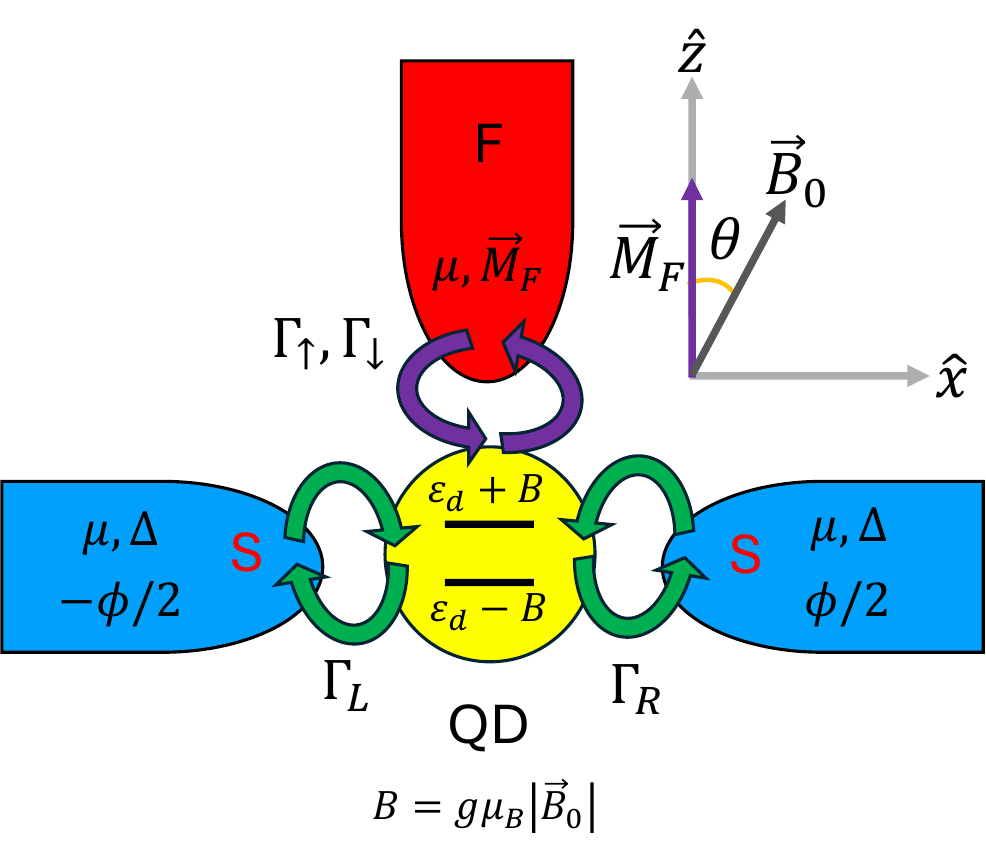}
        		\caption{Scheme of the Quantum Dot Josephson junction connected to the ferromagnetic lead (SQDFS JJ). $\varepsilon_{d}$ is the dot energy level. $\mu$ is the chemical potential of all the leads. $\Delta$ and $\phi$ are the superconducting gap and phase difference between the S leads. The external magnetic field is $\vec{B}_{0}$ and $\vec{M}_{F}$ is the F lead magnetization  with $\theta$ being the angle between the two. The Zeeman splitting on the dot is $B=g\mu_{B}|\vec{B}_{0}|$. 
                $\Gamma$ is the hybridization between the dot and the S leads, $\Gamma_{\uparrow/\downarrow}$ is the spin-dependent coupling to F.} 
		\label{fig: 1_SQDFS_Gamma}
	\end{figure}
    In this section we describe the NH JJ under study. We recall how to compute its complex Andreev levels and the Josephson current within a GF formalism.
    Further details about the model and the techniques can be found in Refs.~\cite{Capecelatro2025, Capecelatro2026}.

    \subsection{Green's function of the quantum dot} \label{sec: model GF}
    We consider a junction made of two superconducting leads (S) and a quantum dot barrier (QD) that is attached to a ferromagnetic metal reservoir (F), see Fig.~\ref{fig: 1_SQDFS_Gamma}.
    The two superconductors have \emph{s-wave} symmetry with gap $\Delta$ and phase-bias $\phi$. The QD is a single energy level $\varepsilon_d$ under an external magnetic field $\vec{B}_{0}$. The Zeeman field $\vec{B}=g\mu_{B}\vec{B}_0$ induces a levels splitting $B=g\mu_{B}|\vec{B}_0|$, with $\mu_{B}$ and $g$ being respectively the Bohr magneton and the gyromagnetic factor. 
    $\vec{B}_0$ can be in general non-collinear with the F lead magnetization, $\vec{M}_{F}$.  We choose $z$ as the spin quantization axis of the ferromagnet and write $\vec{B}=(B_x, 0, B_z)=(B\sin{\theta}, 0, B\cos{\theta})$, $\theta$ being the angle between the two vectors.
    The Hamiltonian of the dot reads 
    \begin{equation}
        \label{H_d}
        H_{d}=\sum_{\sigma=\uparrow,\downarrow}d_{\sigma}^{\dagger}\left(\varepsilon_d+\vec{B}\cdot\vec{\sigma}\right) d_{\sigma},
    \end{equation}
     where $d_{\sigma}$ is the dot annihilation operator for spin $\sigma$ and $\vec{\sigma}=(\sigma_{x},\sigma_{y},\sigma_{z})$ the vector of Pauli operators in the spin space.
    The effect of the S and F leads on the QD can be included via self-energy terms, $\Sigma_{S}=\Sigma_{L}+\Sigma_{R}$ and $\Sigma_{F}$, respectively.
    The dot GF is then defined in Matsubara formalism as $G_{d}\left(\omega_{n}\right)=\left(i\omega_{n}- H_{d}-\Sigma_{S}(\omega_{n}) - \Sigma_{F}(\omega_{n})\right)^{-1}$, where $\omega_{n}=\pi T (2n+1)$ is the $n$-th fermionic Matsubara frequency at temperature $T$.
    
    Since we analyze the equilibrium transport properties of the system, all the leads share the same chemical potential, $\mu=0$, so that no net current flows into the reservoir, which acts as an effective side-coupled reservoir~\cite{Shen2024, Beenakker2024, CayaoSato2024, CayaoSato2024_2, Capecelatro2025, Pino2025, Solow2025, Capecelatro2026}, as clarified in Appendix~\ref{app: GF_current}.
    
    We assume a constant density of states (DOS) in the superconducting leads and symmetric couplings to the dot, $\Gamma_L=\Gamma_R\equiv\Gamma$~\cite{Rozhkov1999, Vecino2003, Sellier2005, Benjamin2007, Karrasch2008, Meng2009_PRB, Zazunov2009, JonckhereeMartin2009, Rodero_Review_2011, Capecelatro2023}.
    Therefore, in Nambu$\otimes$spin space, with $\Psi=\left[d_{\uparrow},d_{\downarrow},-d^{\dagger}_{\downarrow}, d^{\dagger}_{\uparrow}\right]$, the superconducting self-energy reads (see Appendix~\ref{app: QD_GF_and_NH_Hamiltonian})
    \begin{equation}
    \label{S_leads_self_energy}
        \check \Sigma_{S}=\frac{
         \Gamma}{\sqrt{\Delta^2-(i\omega_{n})^2}}\left(i\omega_{n}\check{1}+\Delta\cos{\left(\frac{\phi}{2}\right)}\hat{\tau}_{x}\otimes \hat{1}\right),
    \end{equation}
    where $\hat{\tau}_{x,y,z}$ are the Pauli matrices in Nambu space ($\check{\cdot}$ and $\hat{\cdot}$ denote $4\times4$ and $2\times2$ matrices, respectively).

   In the wide-band limit, the reservoir self-energy is purely imaginary and frequency independent~\cite{Shen2024, Beenakker2024, Capecelatro2025, CayaoSato2024, CayaoSato2024_2, Pino2025, CayaoAguado2024, Solow2025, Capecelatro2026},
    \begin{equation}
    	\label{eq: F_lead_Self_Energy}
        \check{\Sigma}_{F}=-\frac{i\,\mathrm{sign}(\omega_{n})}{2}
        \left[\Gamma_{\uparrow}\left(\check{1}+\hat\tau_{z}\otimes\hat{\sigma}_{z}\right)+
        \Gamma_{\downarrow}\left(\check{1}-\hat\tau_{z}\otimes\hat{\sigma}_{z}\right)\right],
    \end{equation}
    with $\Gamma_{\uparrow,\downarrow}$ the spin-dependent hybridizations between the QD and the reservoir. 
    The difference in dissipation between the two spin channels is a direct consequence of the spin-dependent population of the ferromagnet bands~\cite{CayaoSato2024, Capecelatro2026}. In this picture, having either $\Gamma_{\uparrow}\gg\Gamma_{\downarrow}$ or $\Gamma_{\downarrow}\gg\Gamma_{\uparrow}$ characterizes strongly spin-polarized ferromagnets~\cite{Grein2009, Grein2013, Bobkova2017, Ouassou2017}, while the extreme limit, where only one spin species participates in transport, i.e. either $\Gamma_{\uparrow}\rightarrow0$ or $\Gamma_{\downarrow}\rightarrow0$, corresponds to  half-metal ferromagnets~\cite{deGroot1983, Katsnelson2008, Eschrig2008, Eschrig2015, Keizer2006}. We note that the validity of the wideband approximation for F in our setup is guaranteed as long as the characteristic bandwidth of the ferromagnet is much larger than $\Delta$, which is the relevant energy scale of the JJ, and thus its DOS can be taken constant over the range of the superconducting gap, see Appendix~\ref{app: QD_GF_and_NH_Hamiltonian}. This condition can be fulfilled when considering conventional s-wave electrodes~\cite{James1960, Biondi1957, Wang1996, Pronin1998} and typical strong and half-metal ferromagnets~\cite{Schwarz1986, Soulen1998}.

    Once the dot Green’s function, $\check G_{d}$, is known, both the supercurrent flowing through the junction and the complex Andreev levels can be determined. In particular, as discussed in Sec.~\ref{sec: 2_Josephson_current}, the Josephson current, i.e. the CPR, at finite temperature is directly obtained from the dot Matsubara Green’s function~\cite{Minutillo2021, Capecelatro2023}. 
    Furthermore, upon analytical continuation of $\check{G}{d}(\omega_{n})$ to real frequencies and taking the zero-temperature limit, $i\omega_{n}\rightarrow z=\omega+i\eta$ with $\eta\rightarrow 0^{+}$, one obtains the retarded Green’s function $\check{G}_{d}^{R}(\omega)$~\cite{Capecelatro2023, Capecelatro2025}, see Appendix~\ref{app: QD_GF_and_NH_Hamiltonian}. Its poles define the Andreev levels of the junction, while the local DOS on the dot can be computed accordingly~\cite{Capecelatro2025} (see Sec.~\ref{sec: 0_pi_N_lead}).
   
    \subsection{Josephson Current, non-Hermitian Hamiltonian and Andreev Current}
    \label{sec: 2_Josephson_current}
    The CPR of the junction is directly computed from the GF of the quantum dot $\check{G}_{d}(\omega_n)$ as follows (see Appendix~\ref{app: GF_current})
    \begin{equation}
		\label{Current_Matsubara_Final_chap_4}
		J\left(\phi\right)=  2eT\Delta \Gamma \sin\left(\frac{\phi}{2}\right) \sum_{\omega_{n}}\frac{\mathfrak{R}{\{F_{d}(\omega_{n})\}}}{2\sqrt{\Delta^2+\omega_{n}^2}} \; ,
    \end{equation}
    where $ F_{d}=\left(\check{G}_{d}\right)_{14}-\left(\check{G}_{d}\right)_{23}$ is the \emph{anomalous} part of the dot GF, accounting for the spin-singlet superconducting correlations induced by proximity effect~\cite{Rozhkov1999, Vecino2003, Sellier2005, Benjamin2007, Karrasch2008, Meng2009_PRB, Zazunov2009, JonckhereeMartin2009, Rodero_Review_2011, Capecelatro2023, Capecelatro2025, Capecelatro2026, Furusaki1994, Asano2001, Asano2019, Minutillo2021, Ahmad2022}.
    Eq.~\eqref{Current_Matsubara_Final_chap_4} naturally embodies both the current coming from the GF poles (the Andreev levels) and that from the GF branch-cuts (the continuum states). 
    In the absence of normal metal reservoirs, the continuum contribution can be often disregarded, with the Andreev current being the dominant component of the CPR in many relevant cases, i.e. strong- and weak-coupling limits ($\Gamma\gg\Delta$ and $\Gamma\ll\Delta$).
    
    On the contrary, in junctions coupled to metallic reservoirs, the current carried by the continuum states can get enhanced and be non-negligible. Indeed, the broadening of the levels can lead to a spectral mixing between the sub-gap Andreev resonances and the supra-gap states and the latter can contribute also to the current at energies below $\Delta$~\cite{Capecelatro2025}.
    Nonetheless, in the weak-coupling regime, such contribution remains rather small, compared to the quasi-ABS one, and can often be neglected up to $\Gamma_{N}$ values that are larger than $\Gamma$ but smaller than $\Delta$~\cite{Beenakker2024, Capecelatro2025}.

   For this reason, in this work we focus on this regime where we can study $0-\pi$ transitions by exploiting a simple and transparent NH approach to describe the quasi-ABS. The NH Hamiltonian for this system can be derived from the dot GF $\check G_d$~\cite{Capecelatro2026} and reads (see Appendix~\ref{app: QD_GF_and_NH_Hamiltonian})
   \begin{eqnarray}
        \label{Heff_NH}
        \check{H}_{eff}&=&\varepsilon_{d}\hat{\tau}_{z}\otimes\hat{1}+
        B_{z}\hat{1}\otimes\hat{\sigma}_{z}+B_{x}\hat{1}\otimes\hat{\sigma}_{x}-\nonumber\\
        &&i\Gamma_{N}\check{1}-i\gamma_{N}\hat{\tau}_{z}\otimes\hat{\sigma}_{z}+\Gamma\cos{\left(\frac{\phi}{2}\right)}\hat{\tau}_{x}\otimes \hat{1}\,,
    \end{eqnarray}  
    where we used $\Gamma_{\uparrow}=\Gamma_{N}+\gamma_{N}$ and $\Gamma_{\downarrow}=\Gamma_{N}-\gamma_{N}$.
    The real and the imaginary parts of $\check H_{eff}$ complex eigenvalues represent, respectively, energies and broadenings of the quasi-ABS~\cite{Beenakker2024, CayaoSato2024, CayaoSato2024_2, Capecelatro2025, Pino2025, Solow2025, Capecelatro2026}.
    
    From $\check H_{eff}$ eigenvalues, $z_{j}=\varepsilon_{j}-i\lambda_{j}$, we can compute the Andreev current that, in the zero temperature and infinite gap limit, $T\rightarrow0$ and $\Delta\rightarrow\infty$, reads~\cite{Capecelatro2026} (see Appendix~\ref{app: NH_current})
    \begin{eqnarray}
	   \label{Jpol_T0_simp}
	   J_{\rm ABS}(\phi) \overset{\Delta\rightarrow\infty}{\overset{T\rightarrow0}{=}}J_{\mathrm{Re}}+J_{\mathrm{Im}}&=&-\frac{e}{\pi} \sum_{j} \partial_\phi \varepsilon_j\left( \arctan\left(\frac{\varepsilon_j}{\lambda_j}\right)- \frac{\pi}{2}\right)\nonumber\\&&  -\frac{e}{\pi} \sum_{j}\partial_\phi \lambda_j\ln\left( |z_j|\right).
    \end{eqnarray}

    The two terms $J_{\mathrm{Re}}$ and $J_{\mathrm{Im}}$ describe, respectively, the phase dispersion of the real and imaginary parts of the quasi-ABS~\cite{Capecelatro2025, Capecelatro2026}.
    
    We remark that Eq.~\eqref{Jpol_T0_simp} models exactly the \emph{whole} junction CPR only in the infinite gap limit, i.e. $\Delta\rightarrow\infty$.
    As previously mentioned, in the finite $\Delta$ case, this NH current formula still provides an accurate description of the JJ supercurrent, with the continuum current being much smaller than the Andreev one, despite being finite. 
    
    However, the continuum contribution cannot be disregarded when the Andreev current is very low in amplitude. 
    This is the case of $0-\pi$ transitions at which, as we will see in the next sections, the supercurrent strongly decreases, with the Andreev current becoming as small as the continuum component. This imposes a constraint on the use of Eq.~\eqref{Jpol_T0_simp}. 

    For this reason, we will use the formula in Eq.~\eqref{Current_Matsubara_Final_chap_4} to compute the complete CPR. 
    Nonetheless, Eq.~\eqref{Jpol_T0_simp} will help us in predicting the role of the quasi-ABS in such $0-\pi$ transitions, possibly providing insights on the underlying microscopic mechanisms involved in the process. 

    Finally, we choose to work in the resonant tunneling limit, $\varepsilon_{d}=0$, where analytic expressions for the complex Andreev levels can be obtained~\cite{Capecelatro2026}.
   However, we remark that the following discussion and the results presented below are not strictly limited to this regime, as discussed in Appendix~\ref{app: numerics}.

\section{\label{sec: 0_pi_N_lead} $0-\pi$ transition in NH JJs with normal metal reservoirs}

\begin{figure}[h!]
		\centering
		\includegraphics[scale=0.42]{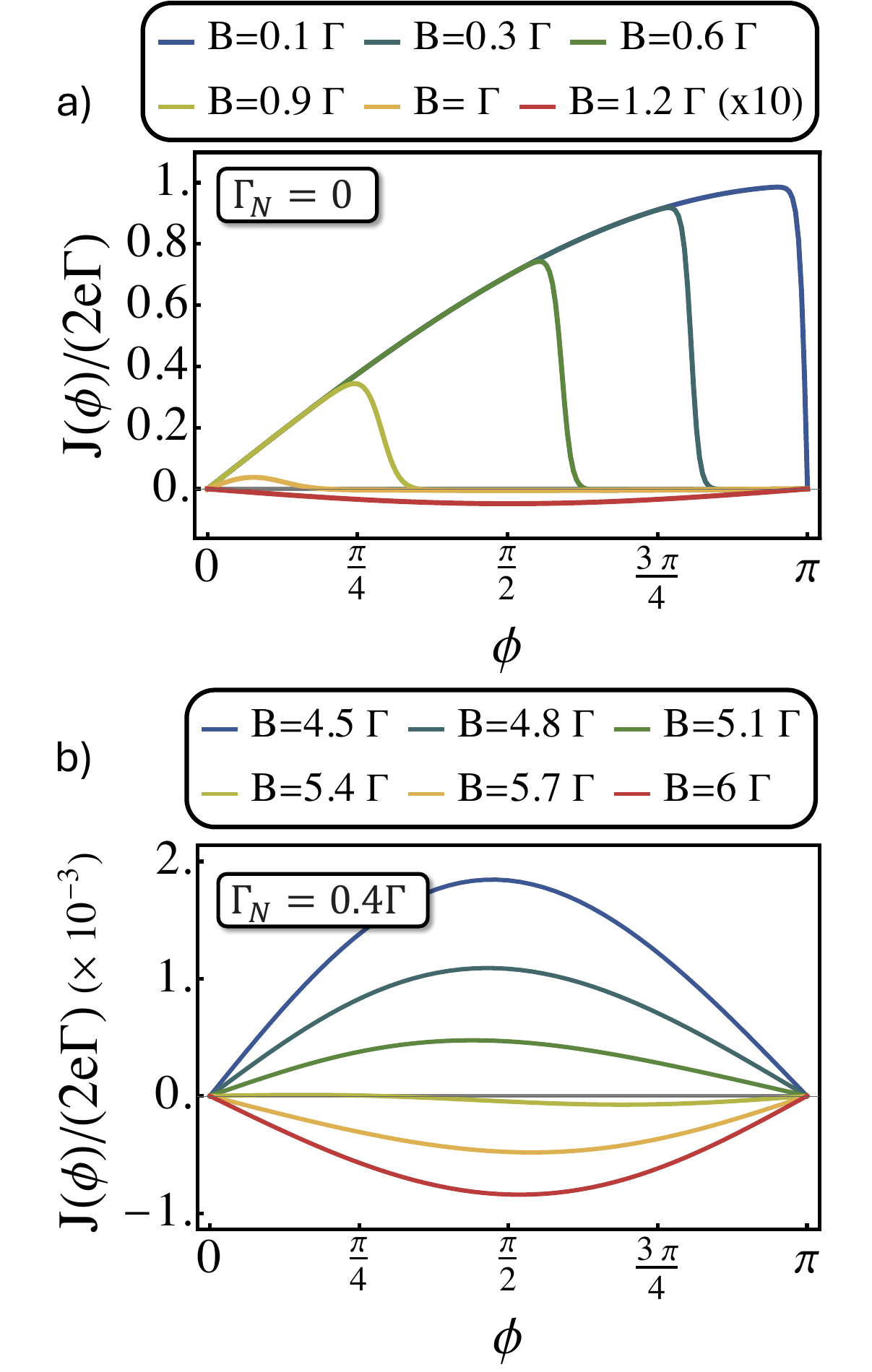}
		\caption{Current-phase relation (CPR) of a Hermitian SQDS JJ (a) and a NH JJ coupled to a N lead (b) along a Zeeman field driven $0-\pi$ transition. As the transition is accomplished the sign of the supercurrent is reversed, thus testifying the phase shift of $\pi$.
        The system parameters are $\varepsilon_{d}=0$, $\Delta=1$ and $\Gamma=0.01$, with $\Gamma_{N}=0$ in (a) and $\Gamma_{N}=0.4\Gamma$ in (b). The temperature is set to $T=2\times10^{-3}T_{C}$.}
		\label{fig: 2_Hermitian_WC_0_pi_transition}
	\end{figure}

 In the absence of the metallic reservoir, the $0-\pi$ transition occurs in such QD JJ when the Zeeman field gets a critical value $B_{0-\pi}$ that, in weak coupling, is equal to the hybridization $\Gamma$ with the two S leads, see Fig.~\ref{fig: 2_Hermitian_WC_0_pi_transition} (a).
    For $B>\Gamma$, the Andreev levels current is completely suppressed while the continuum contribution from the supra-gap states, with negative sign, stays finite, thus setting the junction equilibrium phase to $\pi$~\cite{Benjamin2007, Meng2009_PRB, Capecelatro2023, Capecelatro2025}. 
    The negative continuum current is much smaller than the Andreev contribution~\cite{Capecelatro2023}, typically by two to three orders of magnitude. Nevertheless, because it depends only weakly on the magnetic field~\cite{Benjamin2007, Capecelatro2023}, it becomes increasingly relevant as $B$ is increased and the Andreev current is suppressed, thereby driving the $0-\pi$ transition. 
    
    Further, in the weak-coupling regime the continuum current is largely insensitive to the coupling with the normal lead ($\Gamma_{N}$)~\cite{Capecelatro2025}.
    The difference between Andreev and continuum contributions, under external fields and in the presence of reservoirs, underlies the phenomenology of the $0-\pi$ transitions in NH JJs, as we analyze in the followings.

    When a normal metal lead is attached to the junction, the $0-\pi$ switching shifts to higher magnetic fields, as shown in Fig.~\ref{fig: 2_Hermitian_WC_0_pi_transition}~(b). For $\Gamma_{N}=0$, the transition occurs sharply at $B\gtrsim\Gamma$, with the Andreev current strictly vanishing, see Fig.~\ref{fig: 2_Hermitian_WC_0_pi_transition}~(a). In contrast, for $\Gamma_{N}\neq 0$ the transition occurs more slowly and the CPR evolves smoothly, see Fig.~\ref{fig: 2_Hermitian_WC_0_pi_transition}~(b).
    In this regime, the Andreev contribution, which is strongly suppressed by the coupling to the normal metal, becomes comparable to the continuum one. The interplay between the two contributions, with opposite sign, reshapes the CPR accordingly.
    As discussed below, this behavior results from the competition between the Zeeman splitting $B$ and the coupling to the reservoir $\Gamma_{N}$, and can be understood in terms of the occupation of the quasi-ABS.
    
    For the moment, let us analyze the case of a non-magnetic bath for which $\Gamma_{\uparrow}=\Gamma_{\downarrow}$.
    The shift of the $0-\pi$ transition can be understood when looking at the quasi-ABS of the system that read
    \begin{equation}
        z_{\uparrow,\downarrow}^{e/h}=-i\Gamma_{N}\pm B \pm \Gamma \cos{\left(\frac{\phi}{2}\right)}\,.
    \end{equation}
    Here, the sign in front of $B$ labels the spin sector ($+$ for $\uparrow$, $-$ for $\downarrow$), while the sign in front of $\Gamma \cos(\phi/2)$ labels the \emph{electron-like} ($+$) and \emph{hole-like} ($-$) branches.
    We note that the linewidth is simply $\lambda_{\uparrow/\downarrow}^{e/h}=\Gamma_N$, so that the broadening is phase independent.

    \begin{figure}[h!]
		\centering
		\includegraphics[scale=0.47]{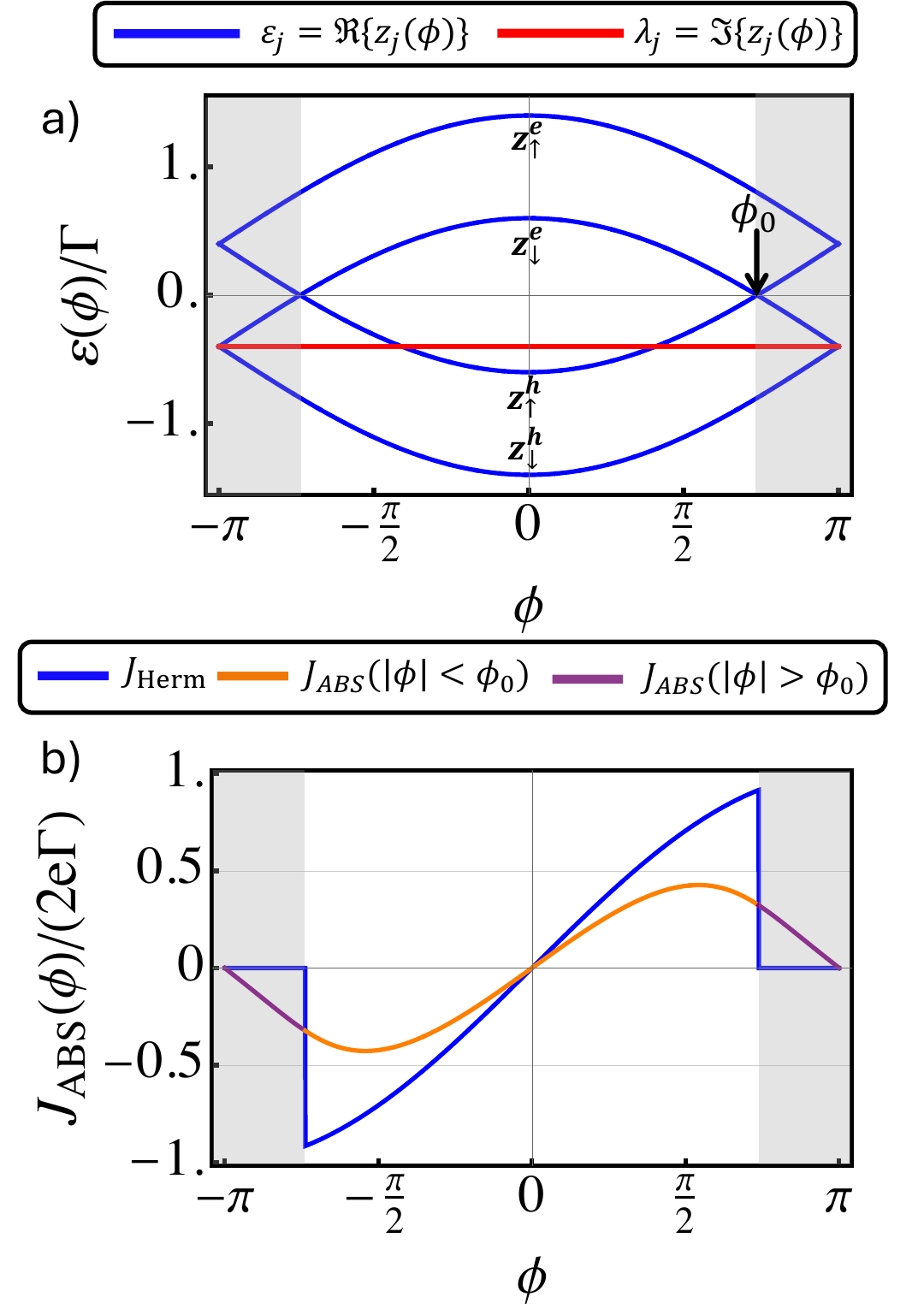}
		\caption{In (a) we show the Andreev levels for the NH JJ with N lead in the presence of a magnetic field. Due to the Zeeman splitting the near-zero energy levels cross at $\phi=\pm\phi_{0}$.
        The system parameters are $\Gamma=1,\,B=0.4,\,\Gamma_{N}=0.4,\,\theta=0$.
        In (b) we compare the corresponding ABS current of the NH JJ with the ABS current of its Hermitian counterpart (at $\Gamma_{N}=0$), at the same $B$. In the Hermitian case outside the ABS crossing (gray shaded regions) the Andreev current vanishes (blue line). In the NH case,  there is a residual Andreev levels current for $|\phi|>\pm\phi_{0}$, even though the critical current is lower.}
		\label{fig: 3_NH_WC_ABS_N_collinear}
	\end{figure}
    The Zeeman splitting leads to the crossing of the spin-down electron channel with the spin-up hole level at $\phi=\pm\phi_{0}=\pm \arccos{\left(\frac{B}{\Gamma}\right)}$, see Fig.~\ref{fig: 3_NH_WC_ABS_N_collinear}~(a).
	As $B$ is increased, the crossing points $\pm\phi_{0}$ move from $\pm\pi$ to $0$, until the spin-up hole and spin-down electron channels are completely switched.
    In a dissipationless junction ($\Gamma_{N}=0$), for $-\phi_{0}<\phi<\phi_{0}$ the current-carrying states are the spin-up hole and spin-down hole levels, i.e. $z_{\downarrow}^{h}$ and $z_{\uparrow}^{h}$, that carry supercurrent in the same direction (the two levels below the chemical potential $\mu=0$ in Fig.~\ref{fig: 3_NH_WC_ABS_N_collinear}~(a)). 
    Outside that interval (gray shaded regions in Fig.~\ref{fig: 3_NH_WC_ABS_N_collinear}~(a) for $|\phi|>\phi_{0}$), the conduction states are the spin-down hole and spin-down electron channels 

    Their supercurrent contributions are equal and opposite so that only the small, negative current carried by continuum states is left over. When the $z_{\uparrow}^{h}$ and $z_{\downarrow}^{e}$ levels get completely switched the system lands finally to the $\pi$ state, see Fig.~\ref{fig: 2_Hermitian_WC_0_pi_transition}~(a). 
    
    When the metallic reservoir is attached to the dot, this mechanism does not hold true anymore.
    In this situation the Andreev current formula in Eq.~\eqref{Jpol_T0_simp} can be rewritten in terms of the current-carrying levels as follows 

    \begin{equation}
    \frac{J_{\rm ABS}}{2e}=- \frac{\partial_\phi \varepsilon^{h}_{\downarrow}}{\pi}\,
    \arctan\!\Bigl(\tfrac{\varepsilon^{h}_{\downarrow}}{\Gamma_N}\Bigr)
    -
    \begin{cases}
    \displaystyle
    \frac{\partial_\phi \varepsilon^{h}_{\uparrow}}{\pi}\,
    \arctan\!\Bigl(\tfrac{\varepsilon^{h}_{\uparrow}}{\Gamma_N}\Bigr),
    & \!\!\!\!\!\! |\phi|\!\!<\!\!\phi_{0},
    \\[1.2ex]
    \displaystyle
    \frac{\partial_\phi \varepsilon^{e}_{\downarrow}}{\pi}\,
    \arctan\!\Bigl(\tfrac{\varepsilon^{e}_{\downarrow}}{\Gamma_N}\Bigr),
    & \!\!\!\!\!\! |\phi|\!\!>\!\!\phi_{0}.
    \end{cases}
    \end{equation}
    
	The $\arctan$ attenuation factors cause the smoothening of the CPR and of the current-jumps typical of the Hermitian system, see Fig.~\ref{fig: 3_NH_WC_ABS_N_collinear}~(b). 
    As a consequence, outside the ABS crossings the cancellation between current-carrying channels is no longer exact, and a finite Andreev contribution survives even for $|\phi|>\phi_0$, see Fig.~\ref{fig: 3_NH_WC_ABS_N_collinear}(b).
    This attenuation is  responsible for the additional robustness of the $0$ state and the delay of the $0-\pi$ transition to $B_{0-\pi}\gg\Gamma$, see Fig.~\ref{fig: 2_Hermitian_WC_0_pi_transition}~(b).
    
    \begin{figure}[h!]
		\centering
		\includegraphics[scale=0.46]{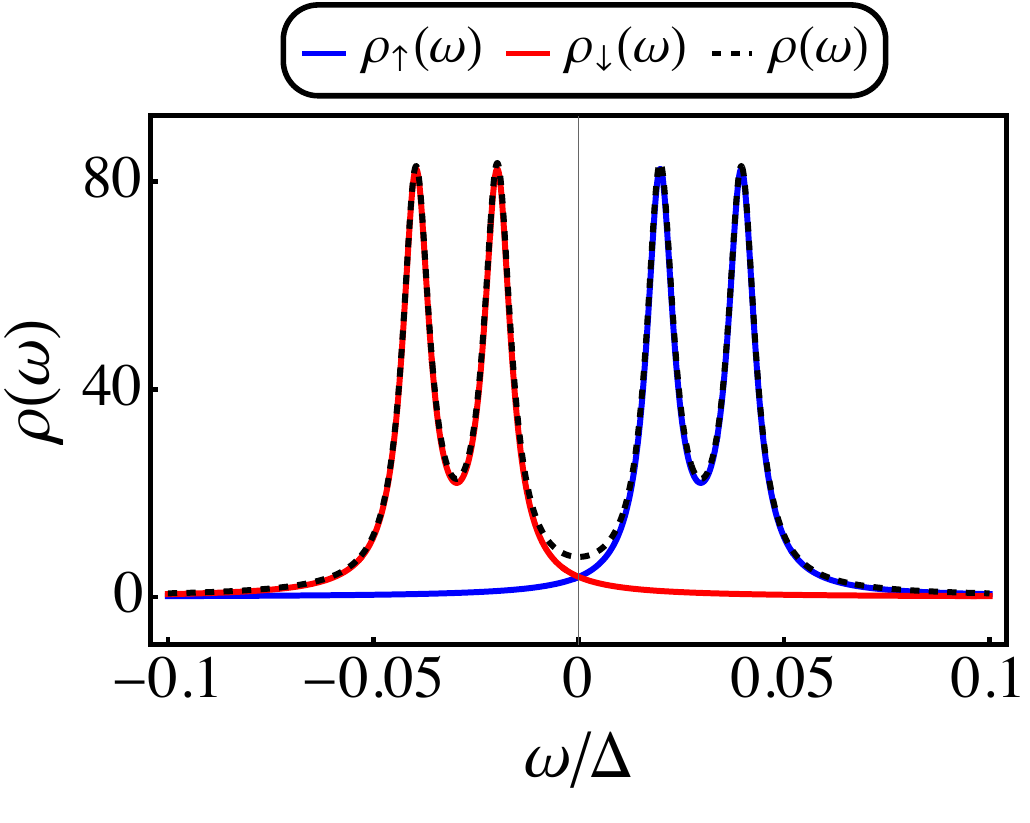}
		\caption{Spin-Resolved QD DOS of a non-Hermitian SQDS JJ. The system parameters are the same as in Fig.~\ref{fig: 3_NH_WC_ABS_N_collinear} except for $B=3\Gamma$.}
		\label{fig: 6b_Panel_Rho_om_NH_JJ_N_lead_SPINRES_0_pi_transition}
	\end{figure}
    
    Physically, this phenomenon can be understood by looking at DOS of the QD, 
    \begin{equation}
      \rho\left(\omega\right)=-(1/\pi)\mathfrak{I}\{\mathrm{Tr}\,\check G_{d}^{R}(\omega)\}  
    \end{equation}
    that can be resolved in the two spin-channels as $\rho\left(\omega\right)=\sum_{\sigma=\uparrow,\downarrow}\rho_{\sigma}\left(\omega\right)=-(1/\pi)\sum_{\sigma}\mathfrak{I}\{\mathrm{Tr}\;\hat G_{d,\sigma}^{R}\}$.
    We report in Fig.~\ref{fig: 6b_Panel_Rho_om_NH_JJ_N_lead_SPINRES_0_pi_transition} the DOS frequency-profile at $B=3\Gamma$. This field is high enough that the aforementioned levels switching is completed, but the $0-\pi$ transition has not been yet accomplished, see Fig.~\ref{fig: 2_Hermitian_WC_0_pi_transition}~(b).
    We note that, due to their finite width, negative-energy states have a finite spectral weight even above the Fermi level. As a result, the two current-carrying levels, with different spectral weight below the Fermi energy, contribute differently to the total current.
    As a consequence, their CPR contributions do not cancel out, leading to a residual ABS current. The $0-\pi$ transition is achieved only for $B\gg\Gamma$, when the spectral weight of both states lies entirely below the Fermi level.

\begin{figure}[h!]
		\centering
		\includegraphics[scale=0.41]{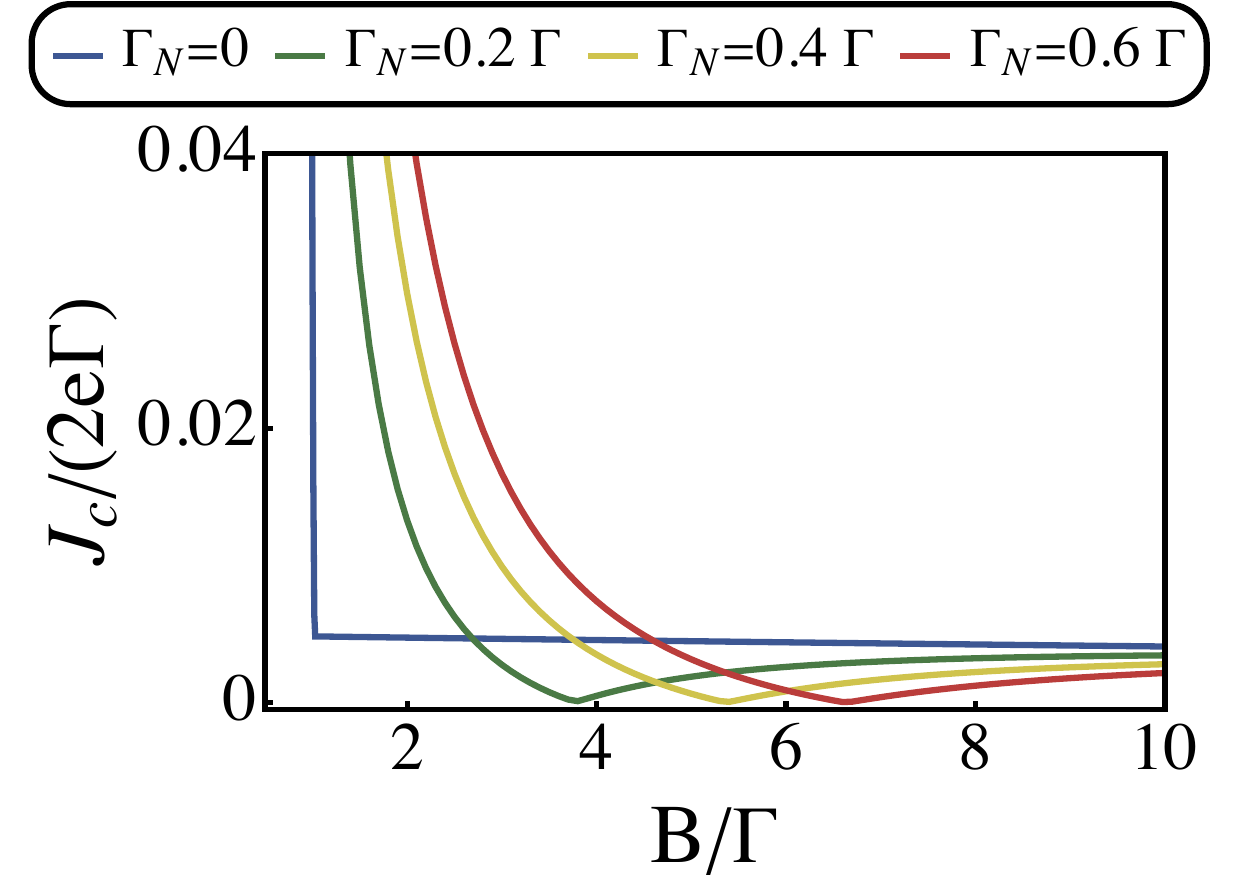}
		\caption{Critical current vs magnetic field amplitude ($J_{c}(B)$) in a NH JJ with a normal metal lead at different values of the coupling $\Gamma_{N}$. The system parameters are the same as in Fig.~\ref{fig: 2_Hermitian_WC_0_pi_transition}~(b) except for $\Gamma_{N}$ and $B$.}
		\label{fig: 5_Jc_vs_B_GammaN_var}
	\end{figure}

    This effect gets amplified when the coupling to the N lead is increased, as we show in Fig.~\ref{fig: 5_Jc_vs_B_GammaN_var}, where the critical current $J_{c}(B)$ is reported for different values of $\Gamma_{N}$. The $0-\pi$ transition fingerprint is the cusp-like behavior that signals a change in the sign of the CPR. In the Hermitian case ($\Gamma_{N}=0$) the transition between the $0$ and $\pi$ states occurs precisely at $B_{0-\pi}=\Gamma$ and its  sharpness reflects the abrupt decrease of the Andreev current. Instead, for $\Gamma_{N}\neq0$ it takes on slowly due to the smoothening of the sawtooth-like behavior shown in Fig.~\ref{fig: 2_Hermitian_WC_0_pi_transition}~(b). Here, the $\pi$ state is reached only when $B\gg\Gamma$.
    
    Interestingly, we can here appreciate the competition between two mechanisms that in principle tend to suppress the Andreev current, the coupling to the environment and the external field.
    The different spectral weights of the current-carrying levels mitigate the current suppression effect of $B$, leading to regimes where $J_{c}$ at high dissipation, e.g. $\Gamma_{N}=0.6\,\Gamma$ curve at $B\sim2\,\Gamma$, is higher than that at much lower dissipation, e.g. $\Gamma_{N}=0.2\,\Gamma$ curve. Finally, by looking at the position of the curve cusps, we can appreciate that the critical field is roughly scaling as $B_{0-\pi} \sim \sqrt{\Gamma_N}$. Such scaling will be analyzed in more detail in App.~\ref{app: asympt_exp}.
    
  \section{$0-\pi$ transition with magnetic reservoirs and tunability with the field angle}
  \label{sec: 0_pi_F_lead}
  The effect of a normal metal lead, inducing a NH term in the JJ Hamiltonian, is a shift in the Zeeman field $B$ of the $0-\pi$ switching.
  Having assessed the different role played by the two different Andreev spin sectors, it is natural to ask what happens to this transition when the junction is coupled to a ferromagnet F with imbalanced tunneling amplitudes $\Gamma_{\uparrow}\neq\Gamma_{\downarrow}$. We find that the interplay of the field $B$ with the spin-dependent couplings can affect the system in different ways.
  \subsection{Dependence of the critical current on the field angle}
  \label{sec: 4_Jc_vs_B_0pi_shift_with_theta}
   \begin{figure}[h!]
		\centering
		\includegraphics[scale=0.41]{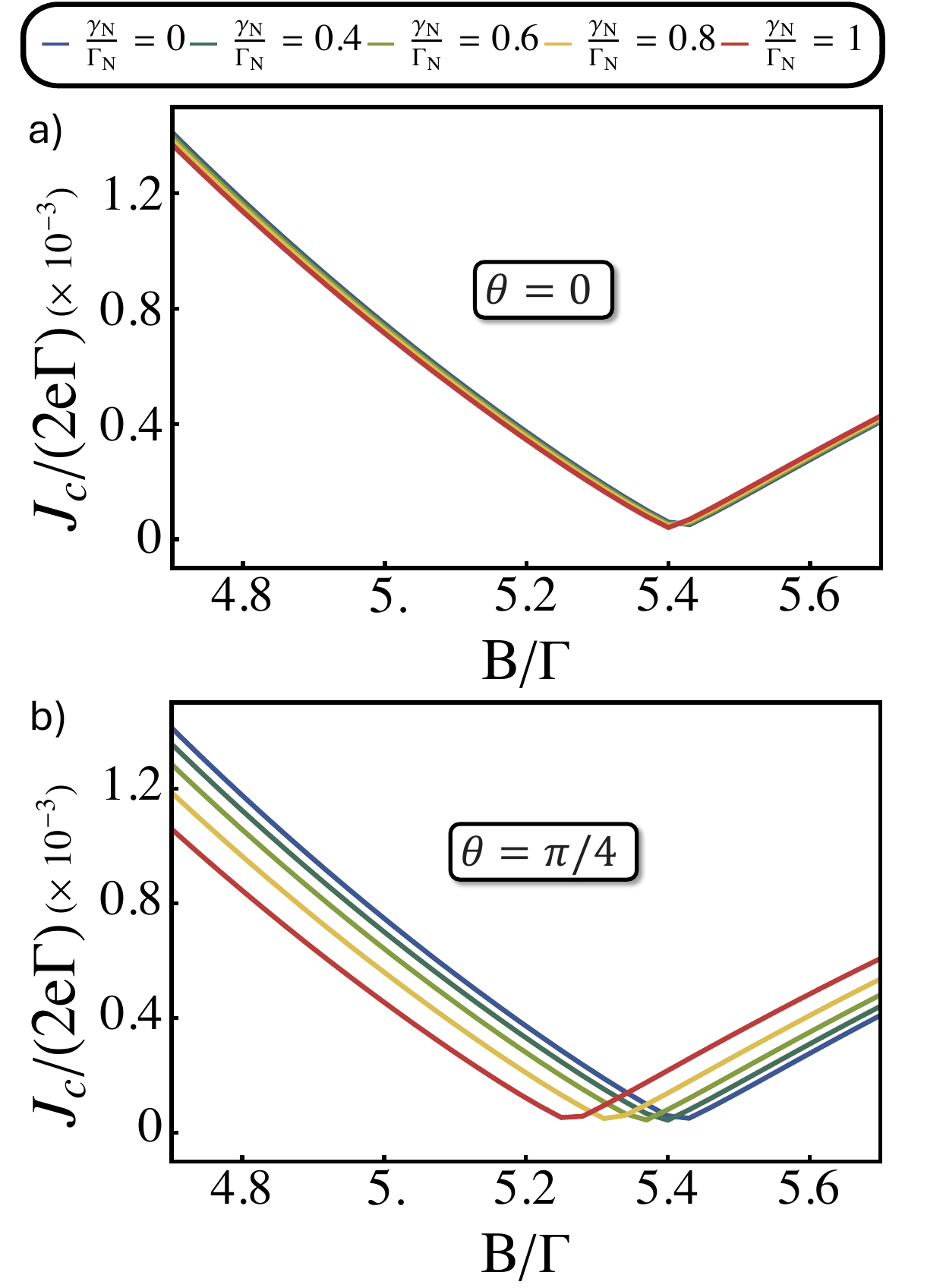}
		\caption{Critical current vs field ($J_{c}(B)$) at different values of the coupling asymmetries between the two spins $\gamma_{N}$ for $\theta=0$ (a) and $\theta=\pi/4$ (b). The other system parameters are the same as in Fig.~\ref{fig: 3_NH_WC_ABS_N_collinear}.}
		\label{fig: 6_Jc_NH_F_lead_gammaNvar_theta_0_Piov4}
	\end{figure}
  
\begin{figure}[h!]
		\centering
		\includegraphics[scale=0.73]{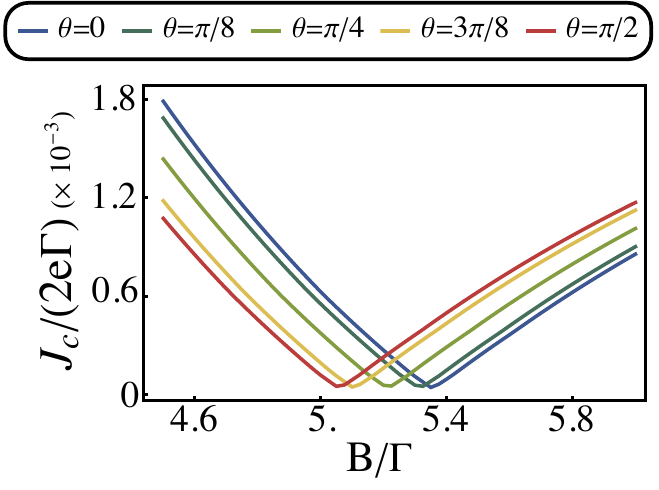}
		\caption{Observing the $0-\pi$ transition from the critical current vs field behavior, $J_{c}(B)$, at different angles $\theta$ between the field and the F lead magnetization.
        The system parameters are $\Delta=1,\,\Gamma=0.01,\,B=4\,\Gamma,\,\Gamma_{N}=0.4\,\Gamma,\,\gamma_{N}=\Gamma_{N}$.}
		\label{fig: 7_Jc_NH_F_lead_non_collinear_B_theta_var}
	\end{figure}

  In Fig.~\ref{fig: 6_Jc_NH_F_lead_gammaNvar_theta_0_Piov4} we report the critical current vs field curves for different values of the dissipation imbalance $\gamma_{N}=(\Gamma_{\uparrow}-\Gamma_{\downarrow})/2$ for $\theta=0$ (a) and $\theta=\pi/4$ (b).
  Surprisingly enough, when the magnetic field is collinear to the F lead magnetization, having a spin-dependent coupling, $\gamma_{N}$, does not affect sizably the $0-\pi$ switching, despite the transition field $B_{0-\pi}$ being slightly lowered when the F lead is completely polarized, i.e. $\gamma_{N}=\Gamma_{N}$. This effect is more visible for $\theta=\pi/4$, see Fig.~\ref{fig: 6_Jc_NH_F_lead_gammaNvar_theta_0_Piov4}~(b), where the typical transition cusp is anticipated by $\sim0.2\,\Gamma$ with respect to the N lead case ($\gamma_{N}=0$). Therefore, when $\vec{B}_{0}$ is not parallel to $\vec{M}_{F}$, the effect of the spin-dissipation imbalance $\gamma_{N}$ on the $0-\pi$ switching is enhanced. More interestingly, the converse is also true. Having a spin-dependent dissipation enables the tuning of the $0-\pi$ transition via the angle $\theta$ between $\vec{B}_{0}$ and $\vec{M}_{F}$. As we can observe in Fig.~\ref{fig: 7_Jc_NH_F_lead_non_collinear_B_theta_var}, the transition field moves to lower values for orthogonal configurations, $\theta=\pi/2$. The angle-tunability of the transition is found also when the QD is slightly detuned from the leads chemical potential, i.e., $\varepsilon_d\lesssim\Gamma$. Therefore, it does not strictly require fine tuning of the QD level at resonance, provided that the junction remains in the high-transparency regime, as discussed in Appendix~\ref{app: numerics}.
	
    We note that the key feature characterizing this shift of the $0-\pi$ transition to lower fields is that $J_{c}$ is a decreasing function of $\theta$. 
   For this reason, we analyze this current behavior by studying the Andreev levels and the Andreev current of the JJ in the next section.
    \subsection{Analysis of the quasi-ABS spectrum}
    The $J_{c}(\theta)$ behavior and the anticipation of the $0-\pi$ transition represent non-trivial effects. We find that they can be explained by analyzing the current contribution from the Andreev levels, $z_{j}=\varepsilon_{j}-i\lambda_{j}$.
    
    They have the form:

    \begin{eqnarray}
		\label{general_ABS_resonance}
        z_{j=1,...,4}&=& - i \Gamma_{N} + s_1\sqrt{
		(B  + s_2\alpha(\theta,\phi)\,)^2 - \gamma_{N}^2 \sin^2(\theta) }\,,\nonumber\\
        &&\alpha(\theta,\phi)=\sqrt{
			  \Gamma^2 \cos^2\left(\frac{\phi}{2}\right)- \gamma_{N}^2 \cos^2(\theta)}\,,
	\end{eqnarray}
    where $j=1,2$ correspond to $s_1=-1$ and $s_2=\pm$, respectively, and $j=3,4$ to $s_1=+1$ and $s_2=\mp$, respectively.
    
\begin{figure}[h!]
		\centering
		\includegraphics[scale=0.45]{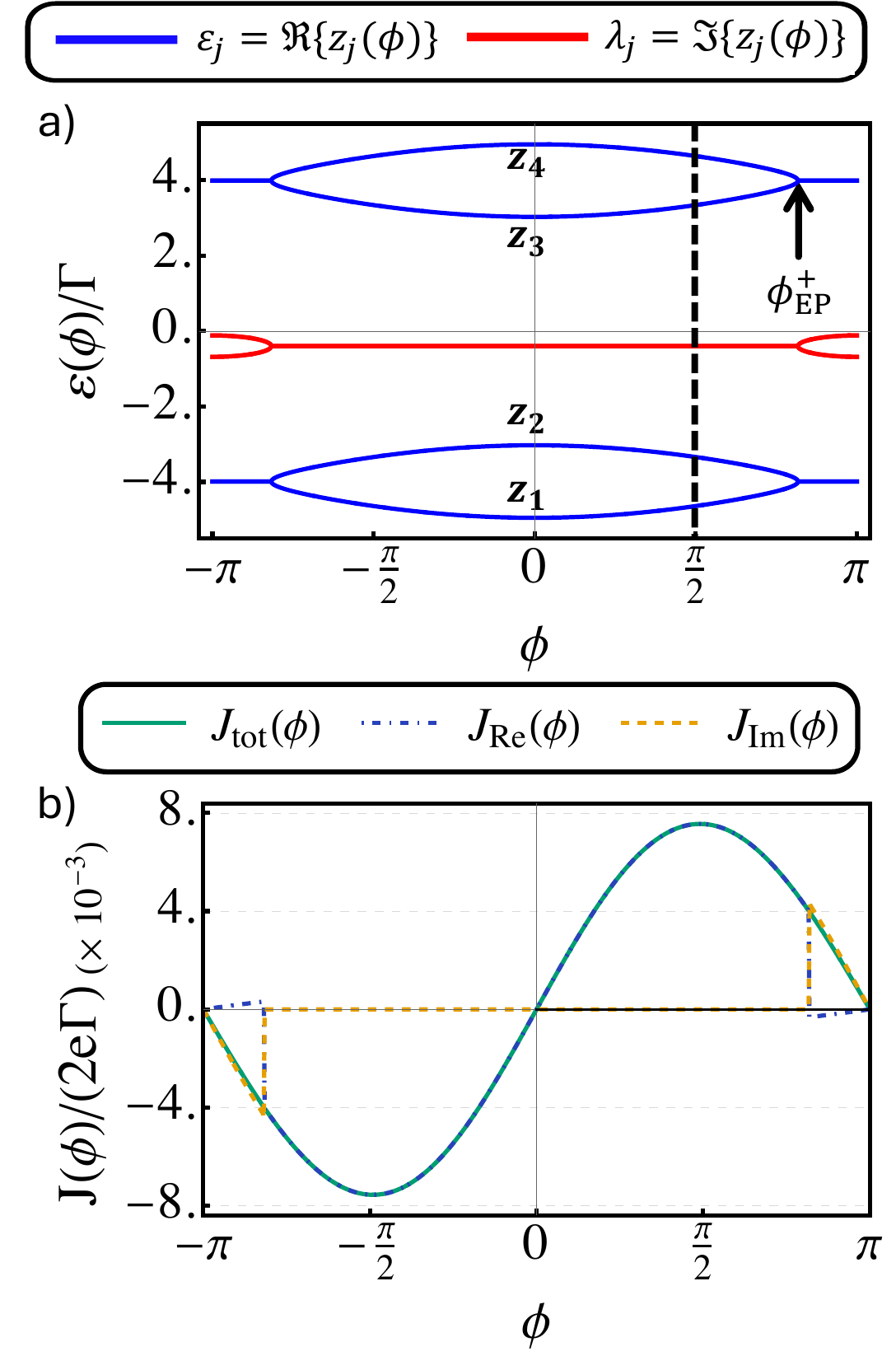}
		\caption{Andreev levels spectrum (a) and corresponding Andreev current (b) 
        for $\vec{B}$ non-collinear with $\vec{M}_{F}$. EPs at finite energy are present. The vertical black dashed line in (a) indicates the energy levels at $\phi=\pi/2$.
        In (b) both the $J_{\mathrm{Re}}$ and $J_{\mathrm{Im}}$ components of the Andreev current are shown.
        The system parameters are $\Gamma=0.01,\,B=4\,\Gamma,\,\Gamma_{N}=0.4\,\Gamma,\,\gamma_{N}=\Gamma_{N},\,\theta=\pi/4$.}
		\label{fig: 8_NH_WC_ABS_F_lead_B_non_collinear}
	\end{figure}

    Such Andreev spectrum can host exceptional points (EPs)~\cite{Solow2025, Capecelatro2026}, which can occur both at zero and finite energy, when the square root term in Eq.~\eqref{general_ABS_resonance}  becomes respectively imaginary or complex.
    The first kind of EPs only originate from the crossings of quasi-ABS at $\varepsilon=0$ and do not occur for $B\gg\Gamma$, for which there are no more levels crossings.
    The second type of EPs are present also for large fields, so that the quasi-ABS spectrum of the system close to transition field ($B\lesssim B_{0-\pi}$) has in general the form of that reported in Fig.~\ref{fig: 8_NH_WC_ABS_F_lead_B_non_collinear}~(a) along with the corresponding Andreev current in Fig.~\ref{fig: 8_NH_WC_ABS_F_lead_B_non_collinear}~(b) computed from Eq.~\eqref{Jpol_T0_simp}.

    In order to understand the current behavior with the tilting angle $\theta$ from the Andreev levels, we analyze $J_{\rm ABS}$ in the limit $B\gg\Gamma,\Gamma_N$. 
    
    For the sake of simplicity, we restrict our computation to a specific phase-bias window. We notice that the imaginary part of the quasi-ABS depends on the phase only after the EPs, as it is evident from Eq.~\eqref{general_ABS_resonance} and Fig.~\ref{fig: 8_NH_WC_ABS_F_lead_B_non_collinear}~(a). The EPs are located at phases 
    \begin{equation} \label{EPs_position}
		\phi_{\text{EP}}^{\pm} = \pm2 \arccos\left( \frac{ \gamma_{N} \cos{(\theta)} }{ \Gamma } \right)\,.
    \end{equation}

    Then, the current-carrying levels in the interval $\phi_{\text{EP}}^{-} < \phi < \phi_{\text{EP}}^{+}$ are simply an imaginary constant plus a real term
    $z_{1,2}=- i \Gamma_{N} - \sqrt{
		(B \pm  \alpha(\theta,\phi)\,)^2 - \gamma_{N}^2 \sin^2(\theta) }$,
    as defined in Eq.~\eqref{general_ABS_resonance}. In this phase-bias window, $J_{\rm ABS}(\phi)$ can be easily computed from  Eq.~\eqref{Jpol_T0_simp}: 
    \begin{eqnarray} \label{currs_pi2}
        J_{\rm ABS}&=&J_{\mathrm{Re}}=J_{z_{1}}+J_{z_{2}}\\  \frac{J_{z_{1,2}}}{(2e)}&=&-\Gamma^{2}\sin(\phi)\;\frac{\pm B +\alpha(\theta,\phi) }{4\pi\alpha(\theta,\phi)\, }\frac{1}{\varepsilon_{1,2}}\;\arctan{\frac{\varepsilon_{1,2}}{\Gamma_{N}}}.
    \end{eqnarray}
    We can now work out an asymptotic expansion of the current contributions for $B \rightarrow \infty$, see App. \ref{app: asympt_exp}, to obtain the total current
    \begin{eqnarray} \label{approx_curr}
        \frac{J_{\rm ABS}}{(2e)}=  \Gamma^{2}\sin(\phi)\left(\frac{\Gamma_N}{2\pi B^{2}} - \frac{\gamma_N^2 s_\theta^{2}}{4B^{3}}\right)  + \mathcal{O}(B^{-4}).
    \end{eqnarray}

    The expression shows that the current vanishes quadratically in $B$, with sinusoidal shape. Moreover, it is entirely due to the coupling with the external lead, $\Gamma_N$. Indeed, in the Hermitian case the asymptotic expansion vanishes since, as soon as $B \geq \Gamma$, the Andreev current vanishes, see Fig.~\ref{fig: 3_NH_WC_ABS_N_collinear}~(b).  
    Concerning the role of the magnetization angle $\theta$, it affects the current at the subleading order $B^{-3}$.
    Notice that for any $\phi$ between the EPs the current decreases by increasing $\theta$ up to $\theta = \pi/2$. The width of this decrease is proportional to the coupling to the external lead, via $\gamma_N$. 

    In the above discussion, we assumed that $\phi_{\text{EP}}^{-} \leq \phi \leq \phi_{\text{EP}}^{+}$ (so that $J_{\rm ABS}\equiv J_{\mathrm{Re}}$).
    It is left to show that also the maximum Andreev current $\mathrm{max}_{\phi}|J_{\rm ABS}|$ is a decreasing function of the tilting angle $\theta$.
    
    This is indeed the quantity that we have to compare to the critical current $J_{c}$ to verify whether the anticipation of the $0-\pi$ transition originates from the quasi-ABS.
    
    We note that the maximum of the asymptotical current function in Eq.~\eqref{approx_curr} is at $\phi=\pi/2$. 
    One can check from Eq.~\eqref{EPs_position}, that such point will always be in the phase-bias interval under examination so long as the coupling with the external lead is not too high, specifically if $\Gamma_{N},\gamma_{N}<\Gamma/\sqrt{2}$.
    As a consequence, our analysis applies to $\mathrm{max}_{\phi}|J_{\rm ABS}|(\theta)$ which, at finite $B$, is well approximated by $J_{\rm ABS}(\phi=\pi/2, \theta)$, under the aforementioned conditions on the dissipation.
    
    To gain physical insight on the phenomenon, one can give a look into the expansion of the single contributions, $J_{z_i}$, which are given in Eq.~\eqref{curr_i_approx} and discussed in App.~\ref{app: asympt_exp}. Phenomenologically, we can say that orthogonal field configurations ($\theta=\pi/2$) slightly hinder the Andreev reflection processes originating the supercurrent with respect to parallel ones ($\theta=0$).
    
    In the following subsection, we utilize $J_{\rm ABS}(\phi=\pi/2, \theta)$ as a benchmark to interpret the supercurrent behavior derived from GF calculations, cf. Eq.~\eqref{Current_Matsubara_Final_chap_4}.

\subsection{The role of continuum states}

The decrease of $J_{\rm ABS}$ with $\theta$ can explain why the $0-\pi$ transition is anticipated when increasing the tilting angle between $\vec{B}_{0}$ and the magnetization in F as follows. As discussed in Sec.~\ref{sec: 0_pi_N_lead}, the sign reversal in the CPR, needed to reach the $\pi$ state, is a result of the competition between the sub-gap current (mainly due to the quasi-ABS) and the supra-gap one.

    The suppression of the Andreev current is stronger for $\theta\neq0$ and thus, under the hypothesis that the supra-gap continuum current (with negative sign) is essentially unchanged, $J(\phi)$ changes its sign at lower values of the field, i.e. $B_{0-\pi}(\theta\neq0)<B_{0-\pi}(\theta=0)$.
    
    To verify this hypothesis and ascribe the whole effect to the NH physics of quasi-ABS, we need to demonstrate that the continuum current is independent from $\theta$. 
    \begin{figure}[h!]
		\centering
		\includegraphics[scale=0.44]{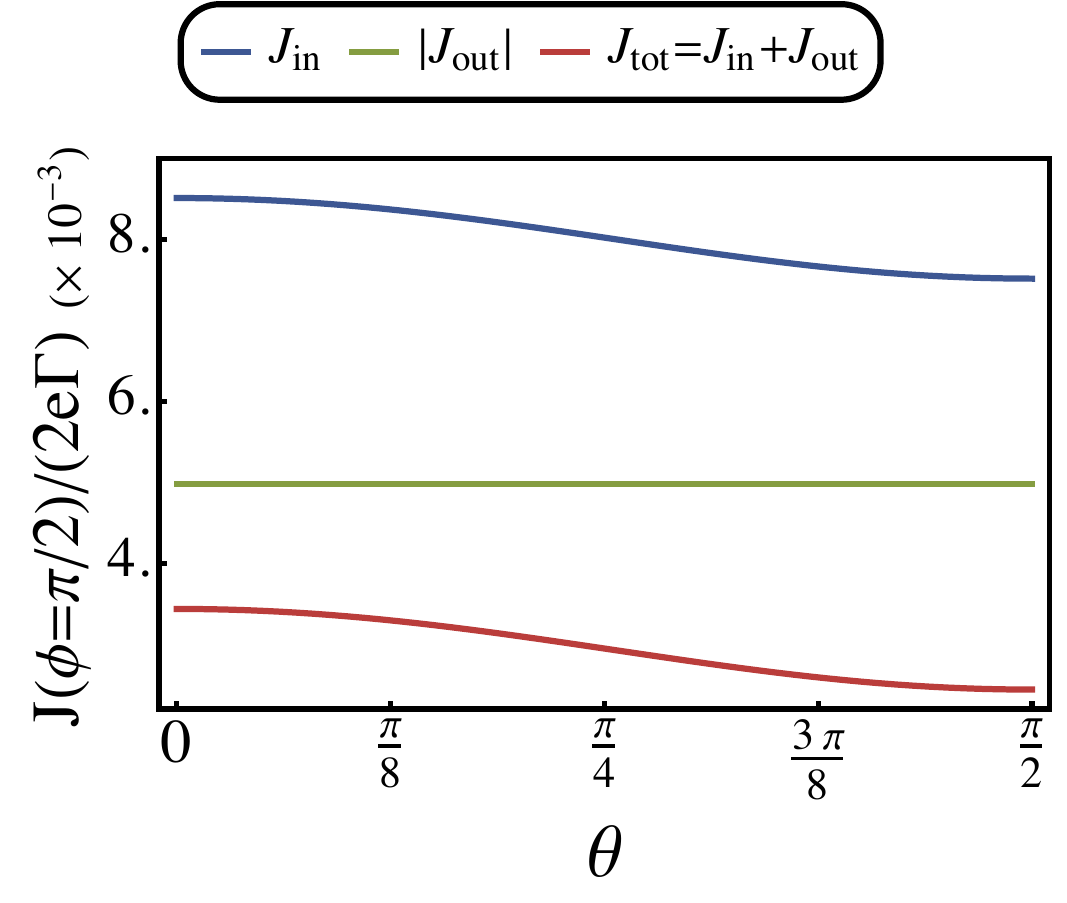}
		\caption{Supercurrent vs magnetic field angle behavior, $J(\phi=\pi/2,\theta)$. The total current $J_{\rm tot}$, the sub-gap current, $J_{\rm in}$, and the absolute value of the negative supra-gap current, $|J_{\rm out}|$,  are reported. The system parameters are the same as in Fig.~\ref{fig: 8_NH_WC_ABS_F_lead_B_non_collinear} except for $\theta$.}
		\label{fig: 9_Jtot_Jin_Jout}
	\end{figure}
    For this purpose, we analyze separately the sub-gap and supra-gap current contributions, $J_{\rm in}$ and $J_{\rm out}$, respectively, which can be derived in the zero-temperature limit from Eq.~\eqref{Current_Matsubara_Final_chap_4} as
    \begin{eqnarray}
		\label{Andreev levels current}
		J_{\rm in}&\overset{T\rightarrow0}{=}&J_{|\omega|<\Delta}=2e\Gamma\Delta\sin{\left(\frac{\phi}{2}\right)}\int_{-\Delta}^{0}\frac{d\omega}{2\pi}\frac{\mathfrak{I}\{F_{d}^{R}\left(\omega\right)\}}{\sqrt{\Delta^2-\omega^2}}\,,\nonumber\\
        J_{\rm out}&\overset{T\rightarrow0}{=}&J_{|\omega|>\Delta}=2e\Gamma\Delta\sin{\left(\frac{\phi}{2}\right)}\int_{-\infty}^{-\Delta}\frac{d\omega}{2\pi}\frac{\mathfrak{R}\{F_{d}^{R}\left(\omega\right)\}}{\sqrt{\omega^2-\Delta^2}}\, ,\nonumber \\
        &&\mathrm{where} \;\;\; J(\phi)\equiv J_{\rm tot}=J_{\rm in}+J_{\rm out}\,.
	\end{eqnarray}
    Results in Fig.~\ref{fig: 9_Jtot_Jin_Jout} confirm that the supra-gap current is insensitive to the magnetic field direction, with the total current profile $J_{\rm tot}(\theta, \pi/2)$ being determined only from the sub-gap component. 
    
    Nevertheless, in open JJs the correspondence between sub-gap and Andreev current does not always hold as mentioned at the end of Sec.~\ref{sec: model GF}. In particular, both the quasi-bound states and the continuum can have a non-negligible spectral weight below the superconducting gap. 
    \begin{figure}[h!]
		\centering
		\includegraphics[scale=0.43]{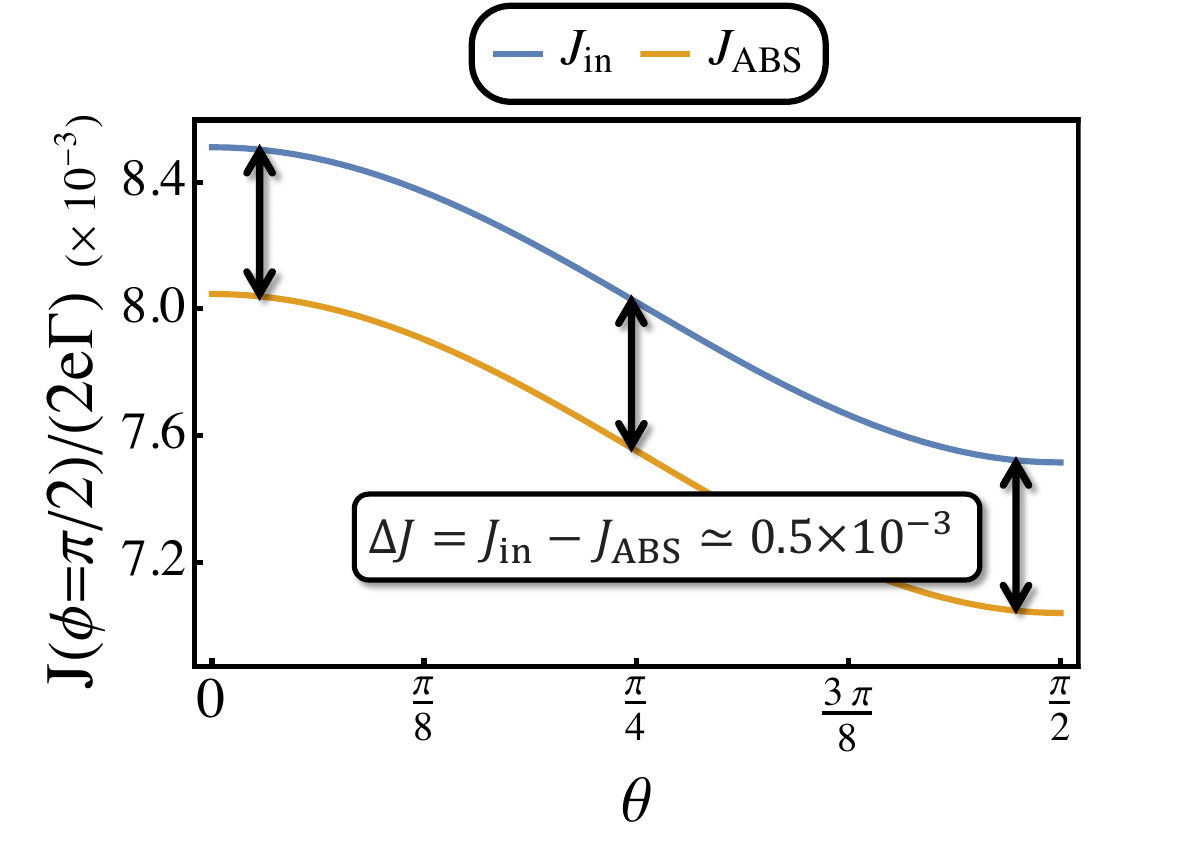}
		\caption{Comparison between the angle behavior of the sub-gap current, $J_{\rm in}(\phi=\pi/2,\theta)$, and that of the Andreev current, $J_{\rm ABS}(\phi=\pi/2,\theta)$. The system parameters are the same as in Fig.~\ref{fig: 8_NH_WC_ABS_F_lead_B_non_collinear} except for $\theta$.}
		\label{fig: 10_Jin_vs_JABS}
	\end{figure}
    Therefore, in Fig.~\ref{fig: 10_Jin_vs_JABS} we compare $J_{\rm in}(\theta, \phi=\pi/2)$ and $J_{\rm ABS}(\theta, \phi=\pi/2)$. 
    
    We find that current modulation with the angle $\theta$ is the same as well as the current decrease between $\theta=0$ and $\theta=\pi/2$.  
    The small difference between the two curves can be ascribed to a small sub-gap continuum current that is more clearly observable when $J_{\rm ABS}$ is strongly dampened~\cite{Capecelatro2025}. 

    Importantly, this component is not affected by the field orientation, analogously to the supra-gap current, thus testifying that the overall $\theta$ dependence of the CPR can be traced back to the quasi-ABS of the JJ.

    The physical reason behind the independence of the continuum currents from the field orientation can be traced back to the fact that supra-gap states have much higher energies (of order $\Delta$) than the magnetic field. Therefore their current will be of tunnel type at the QD, being roughly insensitive to the specific form of $\check{H}_{eff}$ in the weak-coupling limit.

    Our results show that, despite the differences owed to the continuum current contribution, for which $J_{\rm tot}<J_{\rm ABS}$, the Andreev model already captures the relevant physics of the $J(\theta)$ behavior, describing the anticipation of the $0-\pi$ transition with the field angle. We provide a thorough comparison between $J_{\rm tot}(\theta)$ and $J_{\rm ABS}(\theta)$
    in App.~\ref{app: 1_numerical_comparison}. 
    
    As a final remark, we note that the supercurrent is solely carried by singlet correlations, Eq.~\eqref{Current_Matsubara_Final_chap_4}, so that its decrease with $\theta$ can be also understood as a suppression of singlet type correlations in favor of triplet ones. These are promoted by the spin-mixing processes, which occur for $\theta\neq0$ and especially in the orthogonal field configuration ($\theta=\pi/2$), but do not contribute to the CPR, thus resulting in a current reduction.
	Indeed, in the presence of spin-conserving tunneling between the QD and the s-wave spin-singlet superconductors, spin-triplet supercurrents are not allowed, see Appendix~\ref{app: GF_current}.

    \subsection{Tunable $0-\pi$ transition}
    \label{sec: 4_Tunable_0_pi_with_theta}
    \begin{figure}[h!]
		\centering
		\includegraphics[scale=0.44]{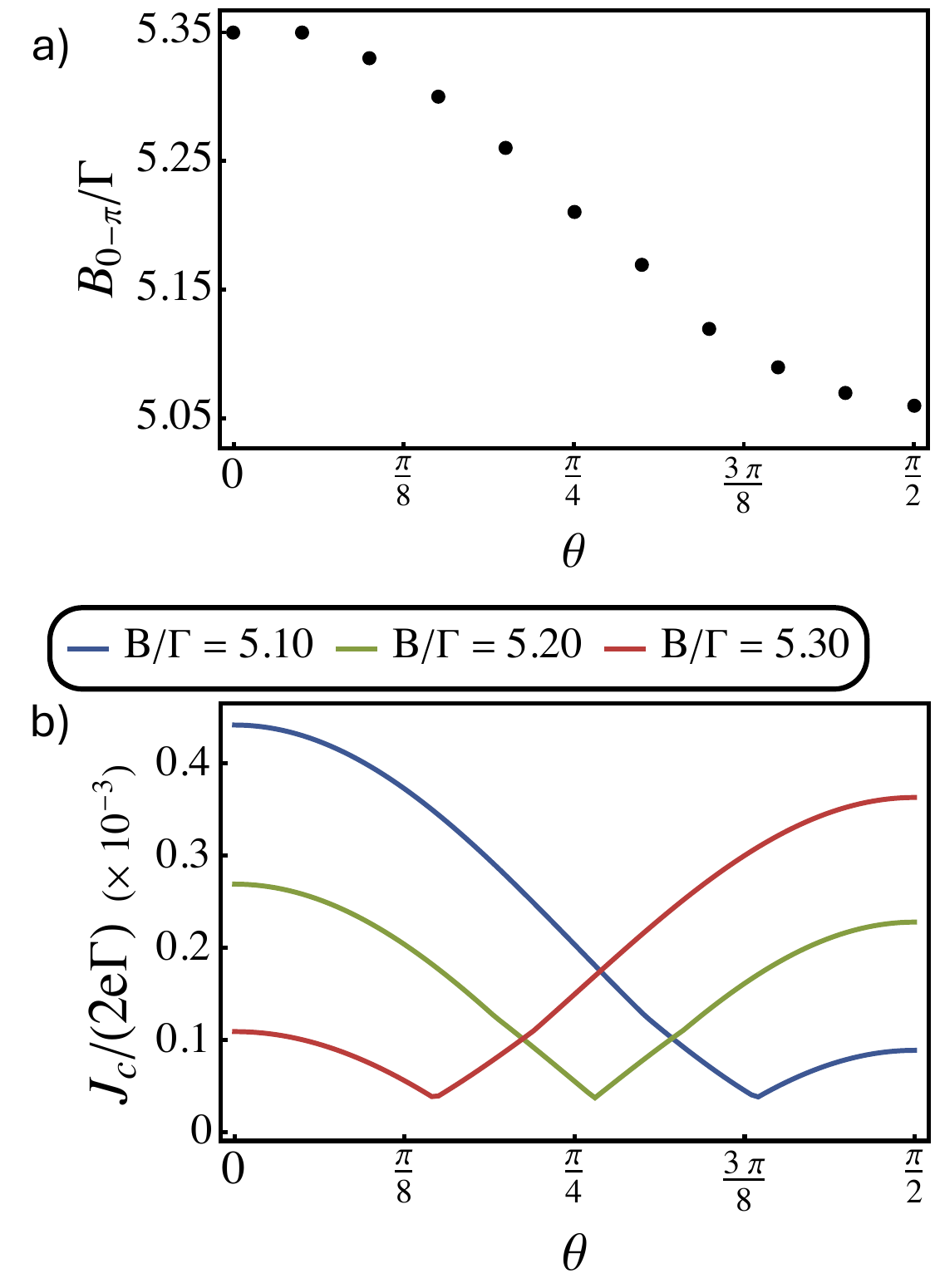}
		\caption{Critical field $B_{0-\pi}$ vs $\theta$ (a) critical current $J_{c}$ vs $\theta$ (b). 
        The system parameters are the same as in Fig.~\ref{fig: 8_NH_WC_ABS_F_lead_B_non_collinear} a part from $B$ and $\theta$.}
		\label{fig: 12_tunable_0_pi_transition}
	\end{figure}
    Our results show how the $0-\pi$ switching can be induced in these simple devices by solely changing the relative angle between $\vec{B}_{0}$ and $\vec{M}_{F}$. 
    The interplay between the spin-dependent decoherence rate and the magnetic field has a non-trivial effect upon the CPR and this phase transition, which is related to the dissipative nature of the coupling between the dot and the reservoir.
    On the one hand, the broadening of the levels moves the $\pi$ phase toward higher-field regimes.
    On the other hand, the non-collinearity between the two magnetic fields does not only introduce spin-mixing and breaks the spin-rotation invariance, lowering the Andreev current, but introduces a specific $J(\phi)$ vs $\theta$ behavior, causing the anticipation of the $0-\pi$ transition. We can, indeed, observe in Fig.~\ref{fig: 12_tunable_0_pi_transition}~(a) how the transition field $B_{0-\pi}$ is lowered when increasing the tilting angle with the F lead magnetization.
    Interestingly enough, this phenomenon might be exploited to tune the equilibrium phase in $0-\pi$ JJs that are coupled to external environments.
    In this respect, in Fig.~\ref{fig: 12_tunable_0_pi_transition}~(b) we show how by keeping the field amplitude fixed and changing its orientation one could switch the JJ phase from $0$ to $\pi$. In particular, by choosing the field intensity $B$ such that $B_{0-\pi}(\pi/2)<B<B_{0-\pi}(0)$ a simple $\sim\pi/2$ rotation of $\theta$ induces the phase switching in the system.
    In practice, we would propose a QD junction coupled to a strongly polarized ferromagnet~\cite{Grein2009, Grein2013, Bobkova2017, Ouassou2017} or to an half-metallic ferromagnet~\cite{deGroot1983, Katsnelson2008, Eschrig2008, Eschrig2015, Keizer2006}, where the intensity of the external field is tuned close to the transition point $B_{0-\pi}(\theta)$ and by changing its orientation with respect to the reservoir magnetization the system moves between the $0$ and $\pi$ states.
        Note that the requirement of a strong spin-polarization in the F lead can be fulfilled by taking for instance half-metallic $\rm{LSMO}$ or $\rm{CrO_2}$ leads where polarizations of order $80\%$ and $90\%$, respectively, can be achieved~\cite{Soulen1998, Schwarz1986}.
    	We also remark that this field-rotation protocol can be fine-tuned by properly choosing the QD heterostructure. The gyromagnetic $g$-factor, which controls the effective external field $B_{0}$ necessary to achieve the $0-\pi$ transition, can strongly vary in different QDs architectures, from the $g\sim2$ in
    	carbon nanotube QDs~\cite{Kuemmeth2008} to $g\sim2-18$ in
    	InAs nanowire QDs~\cite{Bjork2005, dHollosy2013}, up to $g\sim20-70$ in
    	InSb QD heterostructures \cite{Qu2014, Nilsson2009}.
    	As a consequence, the width of the magnetic field window where the $0-\pi$ field-rotation protocol should be applied can be tuned from tens of $\rm{mT}$ down to few $\rm{mT}$, providing a practical knob to optimize the experimental implementation of the protocol.
    	A final comment is deserved for the order of magnitude of the Josephson current close to the $0-\pi$ transition in the weak-coupling regime. We have demonstrated that this limit strictly holds up to $\Gamma/\Delta\sim 1/50$ at which the ABS are still described by the expressions in Eq.~\eqref{general_ABS_resonance}. For instance, for $\Delta\sim2\,\rm{meV}$, as in NbN-based supercoductors, $\Gamma\sim 40\,\mu\rm{eV}$.
    	In this regime, the critical current close to the $0-\pi$ transition can be of the order of tens of pA, which is still experimentally accessible, although rather demanding~\cite{Debbarma2022,Karan2022}.
    	However, we have also tested the robustness of the angle-tuning mechanism up to $\Gamma/\Delta\sim 1/10$, where the larger dot-electrodes coupling leads to critical currents in the range of tens of $\rm{nA}$, see Appendix~\ref{app: numerics}, which is a current scale commonly achieved in superconducting hybrid nanostructures~\cite{Ahmad2022, Trahms2023, Basset2014}.
    	Therefore, with a suitable choice of the QD heterostructure, together with a controlled tuning of the dot-superconductors coupling, the proposed field-rotation protocol can likely meet realistic experimental constraints.

    \section{Summary and outlooks}
    \label{sec: conclusions}
    In this work, we studied the emergence and tuning of the $0-\pi$ transitions in non-Hermitian magnetic Josephson junctions. We focused on a superconductor-quantum dot-superconductor junction where the barrier is coupled to a ferromagnetic metal reservoir in the presence of an external, tilted magnetic field.
    
    We computed the transport properties of the system by using the complex Andreev levels of the junction and benchmarking our results with standard Green's functions calculations.    
    In the Hermitian limit, by increasing the intensity of the field the system can be driven to the $\pi$ state where the equilibrium phase of the junction is $\phi=\pi$. 
    
    The coupling to the environment pushes the phase transition toward higher fields, due to the levels broadening. 
    When a spin-dependent dissipation is considered ($\Gamma_{\uparrow}\neq\Gamma_{\downarrow}$), the angle $\theta$ between the field and the magnetization has an impact on the transition field $B_{0-\pi}$, with the $\pi$ state occurring at lower fields for orthogonal configurations.
    We demonstrate that this behavior can be understood  from the 
    Andreev quasi-bound states, while the continuum current is insensitive to the field orientation.
    The angle dependence of the Andreev levels can be exploited to tune the $0-\pi$ transition by simply changing the field orientation. This represents a different tuning mechanism with respect to Hermitian magnetic junctions. 
    More importantly, this phenomenon shed lights on the role that losses can have in unlocking new tuning parameters for current-phase relation (CPR) engineering. 
    In this respect, an intriguing future perspective would be to investigate the impact of non-Hermiticity on anomalous Josephson effect and Josephson diode effect~\cite{Reynoso2008, Zazunov2009, Yokoyama2013, Yokoyama2014, Campagnano_2015, Minutillo2018, Guarcello2020, Strambini2020, Buzdin2008, Ando2020, Kou2021, Akito2022, Davydova2022, Trahms2023, Nadeem2023, Guarcello2024, Maiellaro2024}.

    In conclusion, our findings demonstrate that $0-\pi$ transitions can occur in magnetic Josephson junctions coupled to electronic reservoirs and that non-Hermiticity provides an additional and conceptually distinct control knob for tunable $0-\pi$ devices. This extends the design space of controllable Josephson elements from closed to open systems and paves the way toward engineered dissipative superconducting devices, with promising implications for superconducting quantum technologies and phase-coherent circuit architectures.

    \begin{acknowledgments}
        This work was supported by PNRR MUR project~PE0000023 - NQSTI (TOPQIN and SPUNTO), by the European Union's Horizon 2020 research and innovation programme under Grant Agreement No~101017733, by the MUR project~CN\_00000013-ICSC, by the  QuantERA II Programme STAQS project that has received funding from the European Union's Horizon 2020 research and innovation programme under Grant Agreement No~101017733, and by PRIN MUR Project TANQU 2022FLSPAJ. 
        This work was partially supported by Horizon Europe EIC Pathfinder under the grant IQARO number 101115190.
        R. Capecelatro, C. Guarcello and R. Citro acknowledge F. Romeo, M. Trama and A. Maiellaro for fruitful discussions. 
    \end{acknowledgments}

\section*{Data Availability Statement}
The data that support the findings of this study are available from the corresponding author upon reasonable request.
\appendix
    \section{QD Green's function and NH Hamiltonian}
\label{app: QD_GF_and_NH_Hamiltonian}
In this section we present the derivation of the effective NH Hamiltonian for the QD, $\check{H}_{eff}$, in Eq.~\eqref{Heff_NH}. 
We start writing a closed Dyson equation for $\check{G}_{d}(i\omega_{n})$ including the dot \emph{bare} GF, $\check{G}_{d}^{0}(i\omega_{n})=(i\omega_{n}-\check{H}_{ d})^{-1}$, and the interactions between the dot and the leads 
\begin{equation}
	\check{G}_{d}(i\omega_{n}) = \left(i\omega_{n}\check{1}-\check{H}_{D}-\sum_{i=L,R,F} 	\check{H}^\dagger_{T_{i}}\check{G}^{0}_{i}(i\omega_{n})\check{H}_{T_{i}}\right)^{-1} \,.
	\label{Gd_Dyson_equation_explicit}
\end{equation}
Here, $\check{G}^{0}_{i}(i\omega_{n})$ is the \emph{bare} GFs of the lead $i=L,R,F$ and $H_{\rm T_{i}}=t_{i}\hat{\tau}_{z}\otimes\hat{1}$ is the tunneling Hamiltonian between the dot and the lead $i$ with amplitude $t_{i}$. 
Note that the above expression is equivalent to 
\begin{equation}
	\check G_{d}\left(\omega_{n}\right)=\left(i\omega_{n}- \check H_{d}-\check \Sigma_{S}(\omega_{n}) - \check \Sigma_{F}(\omega_{n})\right)^{-1}\,,
\end{equation}
when identifying $\check \Sigma_{S}(\omega_{n})=\sum_{i=L,R} 	\check{H}^\dagger_{T_{i}}\check{G}^{0}_{i}(i\omega_{n})\check{H}_{T_{i}}$ and $\check \Sigma_{F}(\omega_{n})=\check{H}^\dagger_{T_{F}}\check{G}^{0}_{F}(i\omega_{n})\check{H}_{T_{F}}$.

For the sake of simplicity we choose the superconductors to be equal, so that $\check{G}^{0}_{R}=\check{G}^{0}_{L}=\check{G}^{0}$. 
$\check{G}^{0}$ has a rather simple form when describing the S leads with a flat band with a constant DOS $\rho_{0}$ at the Fermi energy~\cite{ Zazunov2009,JonckhereeMartin2009,Benjamin2007,Capecelatro2023, Capecelatro2025, Capecelatro2026}, 
\begin{eqnarray}
	\label{S_leads_GF}
	\check G^{0}&=&\frac{
		\pi \rho_{0}}{\sqrt{\Delta^2-(i\omega_{n})^2}}\left[i\omega_{n}\check{1}+\Delta\cos{\left(\frac{\phi}{2}\right)}\hat{\tau}_{x}\otimes \hat{1}\right]\,.
\end{eqnarray}
By choosing symmetric couplings, i.e. $t_{L}=t_{R}=t$ ($\check{H}_{T_{L}}=\check{H}_{T_{R}}$), and introducing the total hybridization parameter with the dot, $\Gamma=2\pi\rho_{0} t^2$, one obtains the S leads self-energy in Eq.~\eqref{S_leads_self_energy}.
The ferromagnet is modeled as a normal 1D semi-infinite chain in tight-binding formalism~\cite{Capecelatro2025, Capecelatro2026, CayaoSato2024}, with chemical potential $\mu_{F}$, hopping amplitude, $\tau_{F}$, and a finite Zeeman splitting, $h$.
Its Green's function reads
\begin{equation}
	\check{G}^{0}_{F}=G_{\uparrow}\frac{\check{1}+\hat{\tau}_{z}\otimes\hat{\sigma}_{z}}{2} + G_{\downarrow}\frac{\check{1}-\hat{\tau}_{z}\otimes\hat{\sigma}_{z}}{2},
\end{equation}
with
\begin{eqnarray}
	G_{\uparrow/\downarrow}\left(i\omega_{n}\right)&=&\frac{\left(i\omega_{n}-(\mu_{F}\pm h)\right)}{2\tau_{F}^{2}}\\
	&&-i\frac{\mathrm{sign}(\omega_{n})}{\tau_F}\sqrt{1-\frac{\left(i\omega_{n}-(\mu_{F}\pm h)\right)^2}{4\tau_{F}^{2}}}\,.\nonumber
\end{eqnarray}
By taking the wideband limit, i.e. $\tau_{F}\gg \omega_{n}$, $G_{\uparrow/\downarrow}$ become 
\begin{equation}
	G_{\uparrow/\downarrow}\left(i\omega_{n}\right)=-i\mathrm{sign}(\omega_{n})\frac{1}{\tau_{F}\pi} \sqrt{1 - \left(\frac{ \mu_{F}\pm h}{2 \tau_{F}}\right)^{2}}\,.
\end{equation}
From the above expression we can compute $\check \Sigma_{F}(\omega_{n})=\check{H}^\dagger_{T_{F}}\check{G}^{0}_{F}(i\omega_{n})\check{H}_{T_{F}}=t_{F}^2\check{G}^{0}_{F}$, whose form is equivalent to that in Eq.~\eqref{eq: F_lead_Self_Energy} when introducing the spin-dependent couplings between the dot and the ferromagnet, i.e. $\Gamma_{\uparrow/\downarrow}=\frac{t_{F}^2}{\tau_{F}\pi}
\sqrt{1 - \left(\frac{ \mu_{F}\pm h}{2 \tau_{F}}\right)^{2}}$.
We observe that the retarded GF of the F lead, $G_{F}^{R}$, can be written in terms of the spin-dependent DOS. Indeed, by definition we have  $\rho_{\uparrow/\downarrow}=-\frac{1}{\pi}\mathfrak{I}\left[G^{R}_{\uparrow\downarrow}(\omega)\right]$, so that 
\begin{eqnarray}
	\label{eq: F_lead_Ret_Self_Energy}
	\check{\Sigma}_{F}&=&-\frac{i}{2}
	\left[\Gamma_{\uparrow}\left(\check{1}+\hat\tau_{z}\otimes\hat{\sigma}_{z}\right)+
	\Gamma_{\downarrow}\left(\check{1}-\hat\tau_{z}\otimes\hat{\sigma}_{z}\right)\right],\\&&\mathrm{with}\;\;\Gamma_{\uparrow/\downarrow}=\pi t_{F}^2\rho_{\uparrow/\downarrow}\,.\nonumber
\end{eqnarray}
This expression of the F lead self-energy clarifies the role of the wideband approximation in our model. This implies taking constant spin DOS $\rho_{\uparrow/\downarrow}$ in F and disregarding retardation effects due to reservoir in the dot GF. In our case, this assumption is justified when the ferromagnet DOS does not show sizable variations over the energy scales of the superconducting gap $\Delta$. Typical $\Delta$ values in conventional BCS superconductors are of the order of $\sim 0.1-1\rm{meV}$~\cite{James1960, Biondi1957, Wang1996, Pronin1998}.
	This scale is much smaller than the characteristic energy scales of metallic reservoirs, whose bandwidths and tight-binding hopping integrals are typically in the eV range~\cite{Mehl1996, harrison2012}.  
	For this reason, the requirements of the wideband approximation can be fulfilled both in strong ferromagnets~\cite{Soulen1998} and in half-metallic ferromagnets~\cite{Schwarz1986}.
	In the case of half-metallic $\rm{CrO_2}$, for instance, the
	DOS varies significantly only at energy scales of $\sim 100\,\rm{meV}$~\cite{Schwarz1986}, thus justifying the use of the wideband description in our setup.

Once we have both the S and F leads self-energies, the retarded dot GF in Nambu$\otimes$spin space reads
\begin{eqnarray}
	\label{Gmatsu}
	\check{G}_{d}(z)& =  \Bigg[& z\left(1+\frac{\Gamma}{\sqrt{\Delta^2-z^2}}\right)\check{1} -\frac{\Delta\Gamma}{\sqrt{\Delta^2-z^2}}\cos{\left(\frac{\phi}{2}\right)}\hat{\tau}_{x}\otimes \hat{1}\nonumber\\
	&&-\varepsilon_{d}\hat{\tau}_{z}\otimes\hat{1}-
	B_{z}\hat{1}\otimes\hat{\sigma}_{z}-B_{x}\hat{1}\otimes\hat{\sigma}_{x}\nonumber\\ 
	&&+i\Gamma_{N}\check{1}+i\gamma_{N}\hat{\tau}_{z}\otimes\hat{\sigma}_{z}
	\Bigg]^{-1} \,,
	\nonumber
\end{eqnarray}
where we use the parametrization $\Gamma_{\uparrow}=\Gamma_{N}+\gamma_{N}$ and $\Gamma_{\downarrow}=\Gamma_{N}-\gamma_{N}$ as in Sec.\ref{sec: Model}.

$\check{G}_{d}(z)$ embodies both the information about the ABS and that about the continuum supra-gap states, respectively encoded in its singular and branch-cut parts 
\begin{equation}
	\label{GF_decomposition}
	\check G_{d}^{R}(z) = \check G^R(z) = \check G^{R}_{pol}(z) + \check G^{R}_{bcut}(z)\,.
\end{equation}

From the knowledge of the GF poles $z_{p}$, the polar GF can be rewritten in terms of a non-Hermitian Hamiltonian matrix $\check H_{eff} = \sum_p z_{p} (R_p \check Z^{-1}\check {P}_{p})$ and its residual matrix $\check Z = \sum_p R_p \check {P}_p$ \cite{Capecelatro2025}
\begin{equation}
	\label{GR_eff_neq}
	\check G^{R}_{pol}(z) = \quad
	\check Z (z - \check H_{eff})^{-1} \,,
\end{equation}
where $R_p\check {P}_p$ is defined as 
\begin{equation}
	R_p\check {P}_p=\frac{1}{2\pi i}\oint_{\mathcal{C}_p} \check G^R(z) dz\,,
\end{equation} 
with $\mathcal{C}_p$ a small contour around $z_p$.

In the weak-coupling regime, that is for large superconducting gap $\Delta\gg\Gamma$, we can simplify the S lead self energy as $\check \Sigma_{S}=\Gamma\cos{\left(\frac{\phi}{2}\right)}\hat{\tau}_{x}\otimes \hat{1}$.
In this case, the retarded dot GF solely coincides with its polar component, $G^{R}(z)=G^{R}_{pol}(z)$, since the supra-gap continuum states lie at much larger energy scales compared to the sub-gap Andreev levels and thus they can be mostly disregarded.
This allows to immediately recognize the frequency-independent NH effective Hamiltonian in Eq.~\eqref{Heff_NH} from the dot GF, i.e. from $\check{G}^{R}(z) \approx \left(z\check{1}-\check{H}_{eff}\right)^{-1}$.
We remark that this is strictly valid only in the infinite gap limit $\Delta\rightarrow\infty$, while in more realistic situations, $\Delta\gg\Gamma$, continuum states contribute to the supercurrent even though they represent the minority current component.
\section{Josephson current from Green's functions}
\label{app: GF_current}
In this appendix we recall how to compute the Josephson current in the Green's function formalism.
At equilibrium conditions, i.e. when all the leads chemical potentials are aligned,
the current flowing from the QD to each lead $i=L,R,F$ can be computed from the GFs of the QD, $\check{G}_{d}$, and of the lead $i$, $\check{G}^{0}_{i}$, as follows~\cite{JonckhereeMartin2009, Capecelatro2025}
\begin{equation}
	\label{Current_Matsubara_text}
	J_{i,d}(\phi)= \frac{i e T}{2}  \sum_{\omega_{n}} \mathrm{Tr} \left[\hat{\tau}_{z}\otimes\hat{1}\left(\hat{H}_{T_{i}}\hat{G}_{d}\hat{H}_{T_{i}}\hat{G}_{i}^{0} -\hat{H}_{T_{i}}\hat{G}_{R}^{0}\hat{H}_{T_{i}}\hat{G}_{d} \right) \right] \, ,
\end{equation}
where $\mathrm{Tr}$ stands for the trace over the Nambu$\otimes$spin space.
We recall that the leads/dot GF matrices in the Nambu space have the following structure 
\begin{equation}
	\label{Nambu_structure_GF}
	\check{G}_{i/d}\left(i\omega_{n}\right)=
	\begin{pmatrix}
		\hat{G}_{i/d} & \hat{F}_{i/d} \\
		\hat{F}_{i/d}^{*} & - \hat{G}_{i/d}^{*}
	\end{pmatrix} \,,
\end{equation}
where $\hat{G}_{i/d}$ and $\hat{F}_{i/d}(i\omega_{n})$ are respectively the normal state and the superconducting correlations in spin space on the leads/dot.
Exploiting the matrix expression in Eq.~\eqref{Nambu_structure_GF} the current flowing between the lead $i$ and the dot, Eq.~\eqref{Current_Matsubara_text}, can be further simplified to
\begin{equation}
	\label{Current_Matsubara_Simp_chap_4}
	J_{i,d} = \dfrac{i e t_{i}^{2}}{2} T \sum_{\omega_{n}} \mathrm{Tr}_{\sigma} \left[\hat{F}_{d} (i\omega_{n}) \hat{F}_{i}^{0,*} (i\omega_{n}) -  \hat{F}_{i}^{0} (i\omega_{n}) \hat{F}_{d}^{*} (i\omega_{n})  \right] \; ,
\end{equation}
where the trace over the Nambu space has been performed and $\mathrm{Tr}_{\sigma}$ is the remaining trace over the spin space. At equilibrium only the superconducting correlation functions of the dot and the bare lead $i$, contribute to the charge current across the junction~\cite{Sellier2005}.
	This explains why at equilibrium the current from the dot to the F lead, $J_{F,d}$ vanishes
	\begin{equation}
		J_{F,d}=-\frac{ieT t_{F}^{2}}{2}\sum_{\omega_{n}}\mathrm{Tr}\left[\hat{G}_{d}\hat{\tau}_{z}\hat{G}_{F}^{0}-\hat{G}_{F}^{0}\hat{\tau}_{z}\hat{G}_{d}\right]=0\,,
	\end{equation}
	since there are no superconducting correlations in F.
When using the expressions of the leads in Eq.~\eqref{S_leads_GF} the supercurrent flowing across the JJ can be cast in the form of Eq.~\eqref{Current_Matsubara_Final_chap_4}.

Equation~(\ref{Current_Matsubara_Simp_chap_4}) also clarifies why the spin-triplet correlation functions generated on the dot for non-collinear magnetic configurations ($\theta\neq0$) do not directly contribute to the Josephson current. In the present system the S leads are conventional s-wave spin-singlet superconductors and the tunneling Hamiltonian is spin-conserving, $\hat{H}_{T_{i=L/R}}=-t_{i=L/R}\hat{\tau}{z}\otimes\hat{1}$. Therefore, the anomalous Green's function of the S lead $i=L,R$, $\hat{F}_{i}^{0}$, can be written as
	\begin{equation}
		\hat{F}_{i}^{0}(i\omega_n)
		=
		f_{s,i}^{0}(i\omega_n)\hat{1}\,.
	\end{equation}
	On the other hand, due to the magnetic structure of the junction, the anomalous Green's function induced on the dot has both singlet and triplet components, thus reading
	\begin{equation}
		\hat{F}_{d}(i\omega_n)
		=
		f_{s,d}(i\omega_n)\hat{1}
		+
		\sum_{\nu=x,y,z}
		f_{t,d}^{\nu}(i\omega_n)\hat{\sigma}_{\nu}.
	\end{equation}
	Substituting these decompositions into Eq.~\eqref{Current_Matsubara_Simp_chap_4}, the spin trace becomes
	\begin{eqnarray}
		&\mathrm{Tr}_{\sigma}
		\left[
		\hat{F}_{d}(i\omega_n)\hat{F}_{i}^{0,*}(i\omega_n)
		-
		\hat{F}_{i}^{0}(i\omega_n)\hat{F}_{d}^{*}(i\omega_n)
		\right]\nonumber
		\\
		&=
		f_{i}^{0,*}(i\omega_n)
		\mathrm{Tr}_{\sigma}
		\left[
		f_{s,d}(i\omega_n)\hat{1}
		+
		\sum_{\nu=x,y,z}
		f_{t,d}^{\nu}(i\omega_n)\hat{\sigma}_{\nu}
		\right]\nonumber
		\\
		&
		-
		f_{i}^{0}(i\omega_n)
		\mathrm{Tr}_{\sigma}
		\left[
		f_{s,d}^{*}(i\omega_n)\hat{1}
		+
		\sum_{\nu=x,y,z}
		f_{t,d}^{\nu,*}(i\omega_n)\hat{\sigma}_{\nu}^{*}
		\right]\nonumber\\
		&=
		2
		\left[
		f_{s,d}(i\omega_n)f_{i}^{0,*}(i\omega_n)
		-
		f_{i}^{0}(i\omega_n)f_{s,d}^{*}(i\omega_n)
		\right] .
	\end{eqnarray}
	All triplet terms are therefore projected out by the spin trace. Hence, in this setup, the Josephson current is sensitive only to the singlet correlations. The triplet components generated on the dot for $\theta\neq0$ can still affect the current indirectly, by modifying the singlet correlations on the dot, but they do not provide a current-carrying channel in this geometry. This is a further consequence of the fact that the ferromagnetic lead is a side-coupled reservoir that is only connected to the QD.

\section{Andreev Current formula in NH junctions}
\label{app: NH_current}
In this appendix, we report a brief derivation summary for the Andreev current formula in Eq.~\eqref{Jpol_T0_simp} that we analyzed in Refs.~\cite{Capecelatro2025,Capecelatro2026}. To this aim, we recall the decomposition of the dot GF in Eq.~\eqref{GF_decomposition} and we make use of the fact that the Josephson current is the phase derivative of  the free energy of the QD JJ~\cite{Benjamin2007, Rozhkov1999}
\begin{equation}
	J(\phi)=-2e\partial_{\phi}\mathcal{F}=-2e T\partial_{\phi}\sum_{\omega_n}\ln\left(\det\left(\check{G}_{d}^{-1}(\omega_n)\right)\right)\,.
\end{equation}
Along this line, we can rewrite the free energy of the GF poles, $\mathcal{F}^{pol}$, and then the Andreev current only for the singular part of the dot GF, $\check{G}^{R}_{pol}$, as a real-frequency integral~\cite{Benjamin2007, Rozhkov1999}
\begin{eqnarray}
	\label{free_energy}
	J_{\rm ABS}(\phi)&=&-2e\partial_{\phi}\mathcal{F}^{pol}= \\
	&=& \frac{-e\partial_{\phi}}{\pi} \int_{-\infty}^{\infty}\mathrm{d}\omega\;   n_F(\omega)   \mathfrak{I} \left\{\ln \det(\left(G^{R}_{pol}\right)^{-1})\right\}\nonumber\\&=&\frac{-e\partial_{\phi}}{\pi} \int_{-\Delta}^{\Delta}  \mathrm{d}\omega\;  n_F(\omega) \mathfrak{I}\left[\prod_j\ln \left(\omega - z_j\right)\right]\nonumber\\ 
	&=&\frac{-e\partial_{\phi}}{\pi} \sum_j\int_{-\Delta}^{\Delta}  \mathrm{d}\omega\;  n_F(\omega) \mathfrak{I}\left[\ln \left(\omega - z_j\right)\right],\nonumber
\end{eqnarray}
where $n_F(\omega)$ is the Fermi distribution function.
This Andreev current accounts solely for the dot GF poles, $z_j=\varepsilon_{j}-i\lambda_{j}$, i.e. the eigenvalues of $\check{H}_{eff}$. 
Performing the zero-temperature and the infinite-gap limits, $T\rightarrow0,\,\Delta\rightarrow\infty$, we obtain the Andreev current in Eq.~\eqref{Jpol_T0_simp}~\cite{Capecelatro2026}.

\section{Other transport regimes and experimental conditions}
\label{app: numerics}
In this appendix, we briefly discuss two transport regimes of interest: the off-resonant and the intermediate-coupling regimes, for which $\varepsilon_d\neq0$ and $\Gamma \lesssim \Delta$, respectively. Although closed analytical expressions for the ABS are not available in these regimes, they are closer to realistic experimental conditions. In particular, we analyze the robustness of the $0-\pi$ transition shift with $\theta$ presented in Sec.~\ref{sec: 4_Jc_vs_B_0pi_shift_with_theta} and of the angular tuning mechanism proposed in Sec.~\ref{sec: 4_Tunable_0_pi_with_theta}.
\subsection{Off-resonance regime}
\label{app: off_resonance}
\begin{figure}[h!]
	\centering
	\includegraphics[scale=0.4]{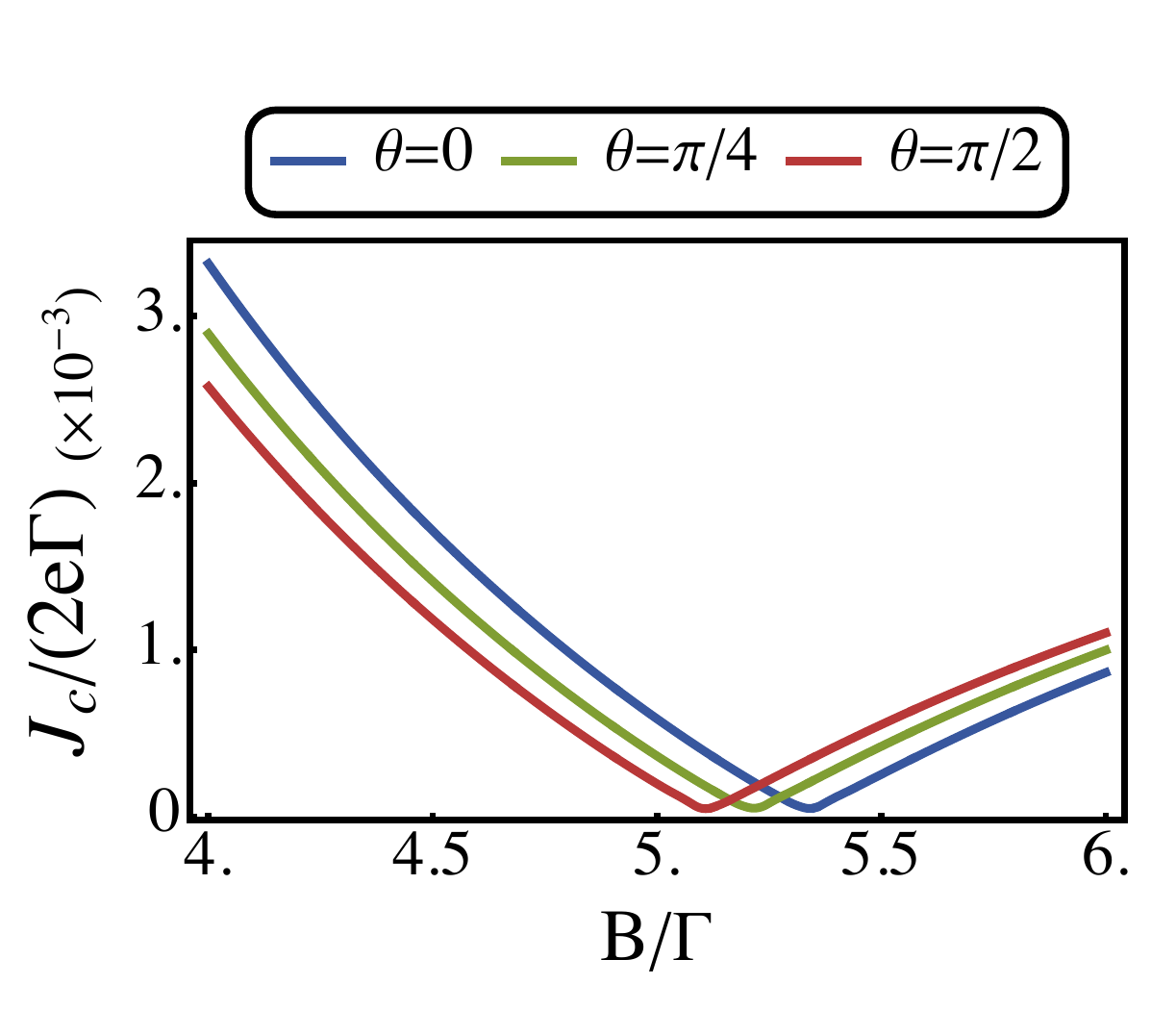}
	\caption{Critical current vs field behavior, $J_{c}(B)$, in the off-resonance regime, $\varepsilon_{d}\neq0$, at different angles $\theta$ between the field and the F lead magnetization.
		The system parameters are $\Delta=1,\,\Gamma=0.01,\,\varepsilon_d=0.2\,\Gamma,\,\Gamma_{N}=0.4\,\Gamma,\,\gamma_{N}=\Gamma_{N}$.}
	\label{fig: 12_Jc_vs_B_eps_0_2_Gamma}
\end{figure}

In this work, we focus on the resonant regime, $\varepsilon_d=\mu=0$, since analytical expressions for the quasi-ABS energies can be obtained only in this limit. 
Nonetheless, in this appendix we test the robustness of the results against small deviations from resonance. In Fig.~\ref{fig: 12_Jc_vs_B_eps_0_2_Gamma}, we show that moderate detunings of the dot level do not qualitatively alter the shift of the $0-\pi$ transition with the angle $\theta$, which is preserved up to dot energy values of $\varepsilon_{d}\sim0.2\Gamma$.
For this reason, the proposed field-rotation protocol for tuning the $0-\pi$ transition is, therefore,  robust for $0<|\varepsilon_d|<\Gamma$.
The phenomenology discussed in the main text is thus preserved, provided that the junction remains in the high-transparency regime. 

\subsection{Beyond the weak-coupling regime}
\begin{figure}[h!]
	\label{app: Gamma_lessim_Delta}
	\centering
	\includegraphics[scale=0.4]{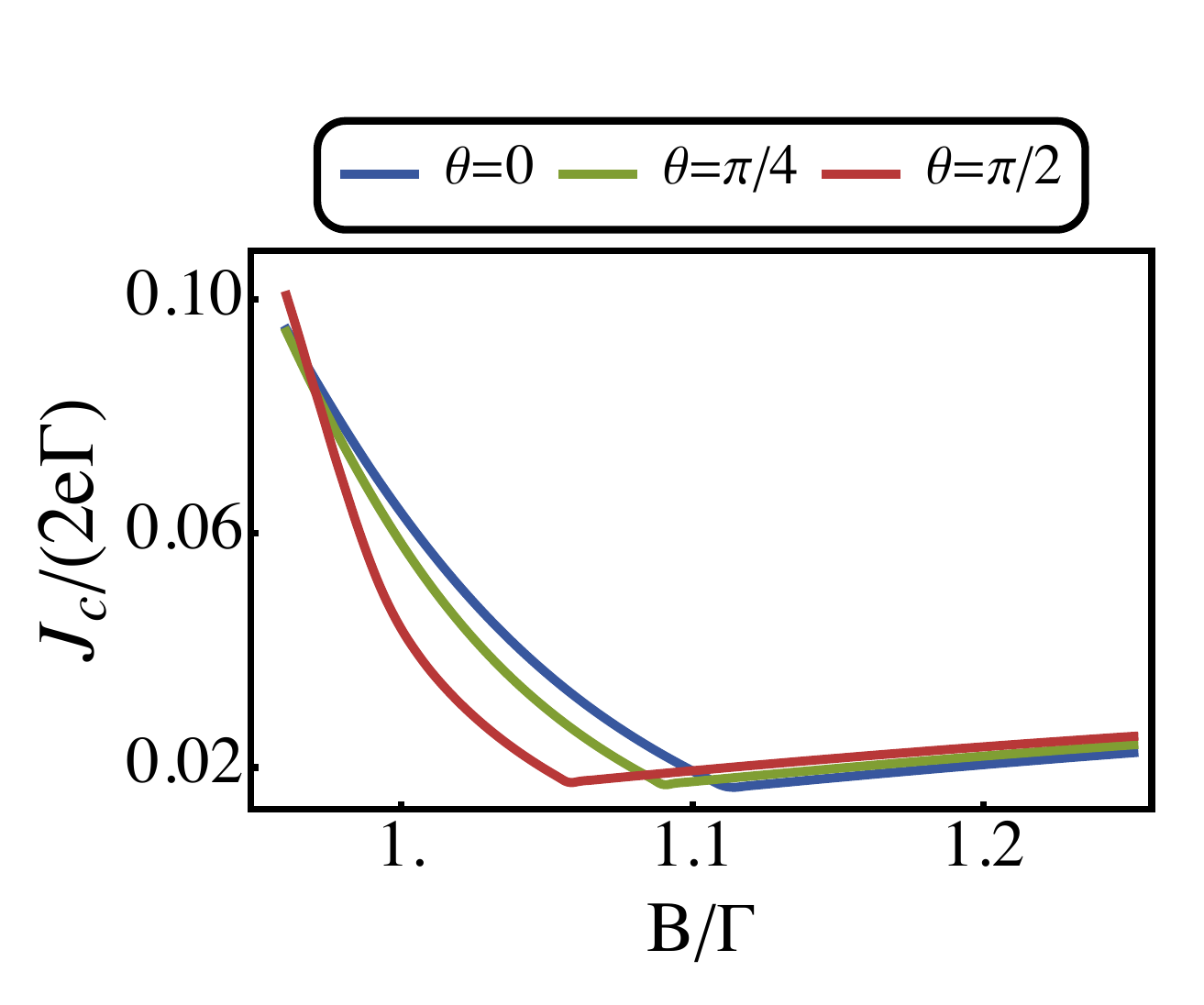}
	\caption{Critical current vs field behavior, $J_{c}(B)$, at different angles $\theta$ between the field and the F lead magnetization.
		The system parameters are $\Delta=1,\,\Gamma=0.1,\,\varepsilon_d=0,\,\Gamma_{N}=0.04\,\Gamma,\,\gamma_{N}=\Gamma_{N}$.}
	\label{fig: 13_Jc_vs_B_Gamma_0_1_Delta}
\end{figure}

In this appendix, we study the shift of the $0-\pi$ transition with the field angle $\theta$ in QD JJs with $\Gamma\lesssim\Delta$. In this transport regime, the supercurrent that can flow through the JJ is much higher than in the weak-coupling limit, thus, potentially making the experimental detection simpler.
We can indeed observe in Fig.~\ref{fig: 13_Jc_vs_B_Gamma_0_1_Delta} that the angle-tuning of the $0-\pi$ transition can be still performed also in a JJ with  $\Gamma=0.1\,\Delta$.
Here, this effect generally persists at lower dissipation, e.g. $\Gamma_{N}=0.04\,\Gamma$, thus favoring higher Andreev currents.

	This transport regime, thus, enables a simpler application of the field-rotation protocol for the $0-\pi$ transition. 
	For instance, considering Nb- or NbN-based superconducting electrodes with a gap of order $\Delta\sim 2\,\mathrm{meV}$~\cite{Wang1996,Pronin1998}, the choice $\Gamma=0.1\Delta$ corresponds to a dot-lead coupling $\Gamma\sim 200\,\mu\mathrm{eV}$.
	This sets a natural current scale $2e\Gamma/\hbar$ of order $10^2\,\mathrm{nA}$. Therefore, normalized critical currents in the range $J_c/(2e\Gamma)\sim 10^{-2}-10^{-1}$ correspond to physical currents of order $1-10\,\mathrm{nA}$,
	which are within the experimental current range achievable in superconducting hybrid nanostructures~\cite{Basset2014,Trahms2023, Ahmad2022}.
	Although increasing $\Gamma$ pushes the $0-\pi$ transition field $B_{0-\pi}$ to larger values, sizable Zeeman splittings can be achieved in semiconductor QDs with large effective $g$-factors. For instance, $g$-factors of order $g\sim 2-18$ have been reported in InAs QDs~\cite{Bjork2005,dHollosy2013}, while values $g\sim 20-70$ can be reached in InSb QDs~\cite{Qu2014,Nilsson2009}.
	Considering InAs nanowire QDs with $g\sim15$, the Zeeman splitting necessary for the transition $\sim 200\,\rm{\mu eV}$ can be achieved with external magnetic fields $B_{0}\sim 230 \,\rm{mT}$, which can be still sustainable in Nb-based heterostructures~\cite{Gul2017, Joshi2017, Wei2021}. 
	In this case, the width of the $0-\pi$ transition window in which the field-rotation protocol should be applied is of order $40\,\mathrm{mT}$, which is experimentally accessible. 
	$0-\pi$ transition fields of the order of tens of $\rm{mT}$ can be reached in InSb QDs with $g\sim 30-40$.
	Further, by controlling the $\Gamma/\Delta$ ratio one can also tune the $0-\pi$ magnetic field interval from hundreds of $\rm{mT}$ to few $\rm{mT}$.
	These findings suggest that with a suitable choice of the QD heterostructure, together with a proper control of the coupling to the S leads, the field-rotation protocol can met experimentally accessible conditions.

\section{Expansion of the Andreev currents at high B}
    \label{app: asympt_exp}
    We show here the asymptotic expansion in $B\rightarrow \infty$ of the current contribution $J_{z_j}$ appearing in Eq.~\eqref{currs_pi2}, up to third order. We start expanding separately the energy derivative and the occupation factor (we restrict to those relative to the $z_1$ level):
    \begin{eqnarray}
    	\label{pieces_approx}
    \partial_\phi \epsilon_1 &=& -\frac{\sin(\phi)\Gamma^{2}(B+\alpha)}{4\alpha\epsilon_{1}}
    =\\
    &&
    \frac{\sin(\phi)\Gamma^{2}}{4\alpha} \left(\!1\!
    +\!\frac{\gamma_N^{2}s_{\theta}^{2}}{2\,B^{2}} \!
    -\!\frac{\alpha\gamma_N^{2}s_{\theta}^{2}}{B^{3}}\right),
    \nonumber\\[6pt]
    \arctan\!\left(\frac{\epsilon_{1}}{\Gamma_N}\right)
    &=&
    -\frac{\pi}{2}
    +\frac{\Gamma_N}{B}
    -\frac{\Gamma_N\alpha}{B^{2}} +\nonumber\\
	&&\quad\frac{6\Gamma_N\alpha^{2}+3\Gamma_N\gamma_N^{2}s_{\theta}^{2}-2\Gamma_N^{3}}{6B^{3}},
\end{eqnarray}

where we shortened $\alpha(\theta,\phi) \rightarrow \alpha$, $\sin\theta \rightarrow s_\theta$. 
Therefore, for the current contributions we have (we assume $\phi_{\text{EP}}^{-} < \phi < \phi_{\text{EP}}^{+}$)
\begin{eqnarray} \label{curr_i_approx}
		\frac{J_{z_1}}{(2e\sin(\phi))}&=& -\frac{\Gamma^{2}}{4\pi\alpha}
		\frac{B+\alpha}{\epsilon_{1}}
		\arctan\!\left(\frac{\epsilon_{1}}{\Gamma_N}\right) \\
		&\sim&
		\frac{\Gamma^{2}}{8\alpha}
		-\frac{\Gamma^{2}\Gamma_N}{4\pi\alpha\,B}
		+\Gamma^{2}\left(
		\frac{\Gamma_N}{4\pi}
		+\frac{ \gamma_N^{2}s_{\theta}^{2}}{16\alpha}
		\right)\frac{1}{B^{2} 
		} +\nonumber\\
		&&
		\Gamma^{2}\left(
		\frac{3\Gamma_N(2 \alpha^2 + \gamma_N^2 s_\theta^{2}) -  3 \pi \alpha \gamma_N^2 s_\theta^{2}  - 2 \Gamma_N^3}{24\pi\alpha\,B^{3}}
		\right)\nonumber\,.
\end{eqnarray}

The current contribution $J_{z_2}$ is obtained by simply sending $\alpha$ to $-\alpha$. 
	\begin{eqnarray}
		J_{z_2} &=& J_{z_1}\vert_{-\alpha}\,,\\
		\frac{J_{z_2}}{(2e\sin(\phi))} &\sim&-\frac{\Gamma^{2}}{8\alpha}
		+\frac{\Gamma^{2}\Gamma_N}{4\pi\alpha\,B}
		+\Gamma^{2}\left(
		\frac{\Gamma_N}{4\pi}
		-\frac{ \gamma_N^{2}s_{\theta}^{2}}{16\alpha}
		\right)\frac{1}{B^{2} 
		} +\nonumber\\
		&&
		\Gamma^{2}\left(
		\frac{-3\Gamma_N(2 \alpha^2 + \gamma_N^2 s_\theta^{2}) -  3 \pi \alpha \gamma_N^2 s_\theta^{2}  + 2 \Gamma_N^3}{24\pi\alpha\,B^{3}}
		\right)\nonumber.
	\end{eqnarray}
	The above $B\rightarrow\infty$ expressions originate from the fact that $z_{1}$ and $z_{2}$ carry current contributions with opposite sign that are weighted by different broadening factors. The overall angle-dependence of $J_{z_{1}}$ and $J_{z_{2}}$ is provided by the factor $\alpha(\theta,\pi/2)^{-1}$, which is a decreasing function of $\theta$ for $\theta\in[0,\pi/2]$.
When computing $J_{\mathrm{ABS}}=J_{z_{1}}+J_{z_{2}}$ from the expansion in Eq.~\eqref{curr_i_approx}, the even terms in $\alpha(\theta)$ are those that will contribute to the total current in summation.
The leading orders are equal and opposite up to $\mathcal{O}(B^{-1})$ terms. It follows that the total current will scale like $B^{-2}$.
The $\theta$ dependence will come from the $\mathcal{O}(B^{-3})$ terms.
The total current is 
\begin{eqnarray}
	\frac{J_{\mathrm{ABS}}}{(2e)}=  \Gamma^{2}\sin(\phi)\left(\frac{\Gamma_N}{2\pi B^{2}} - \frac{\gamma_N^2 s_\theta^{2}}{4B^{3}}\right)  + \mathcal{O}(B^{-4}),\end{eqnarray}
which is Eq.~\eqref{approx_curr}. 

To gain further physical insight into this result, we utilize the expansion in Eq.~\eqref{pieces_approx}. The $\mathcal{O}(B^{-2})$ terms in the total current comes ultimately from the occupation factor. The $\theta$ dependence, however, is provided entirely by the energy derivative term via its $\mathcal{O}(B^{-3})$ term.

We note that this formula may help also in understanding the square-root dependence of $B_{0-\pi}$ from $\Gamma_{N}$. Indeed, since the sub-gap and the supra-gap currents, respectively $J_{\mathrm{in}}$ and $J_{\mathrm{out}}$, have opposite signs, $B_{0-\pi}$ can be defined as the field at which they get equal in magnitude. Since $J_{\mathrm{in}} \sim J_{\mathrm{ABS}}$ and $J_{\mathrm{out}}$ is roughly independent of $B$, by keeping only the leading order in Eq.~\eqref{approx_curr}, at the $0-\pi$ transition we have
$J_{\mathrm{ABS}}/2e\simeq \Gamma^{2}\sin(\phi)\left(\frac{\Gamma_N}{2\pi B^{2}}\right)\simeq |J_{\mathrm{out}}|/2e$. It follows that $B_{0-\pi}   \propto \sqrt{\Gamma_N}$.
    \section{Comparison between full Green's function calculation and the Andreev NH model}
    \label{app: 1_numerical_comparison}
    In this appendix, we provide a numerical comparison between the full GF calculation (Eq.~\eqref{Current_Matsubara_Final_chap_4}) of $J(\theta)$ behavior and that predicted with the Andreev NH model (Eq.~\eqref{Jpol_T0_simp}). Our goal is to quantify the quality of the Andreev description in predicting the system transport properties.
    
    We refer to the L1-norm of the current $\eta$ defined as
    
    \begin{eqnarray}
    \label{eq: eta_def}
     \eta
     &=& \int_{0}^{\pi} d\phi\,\big|J(\phi)\big|\nonumber,
    \end{eqnarray}
    
    where the integral can be computed for $\phi\in\left[0,\pi\right]$ due to the symmetry properties of the CPR, i.e. $J(\phi)=-J(-\phi)$ and thus $|J(\phi)|=|J(-\phi)|.$    
    \begin{figure}[h!]
		\centering
		\includegraphics[scale=0.43]{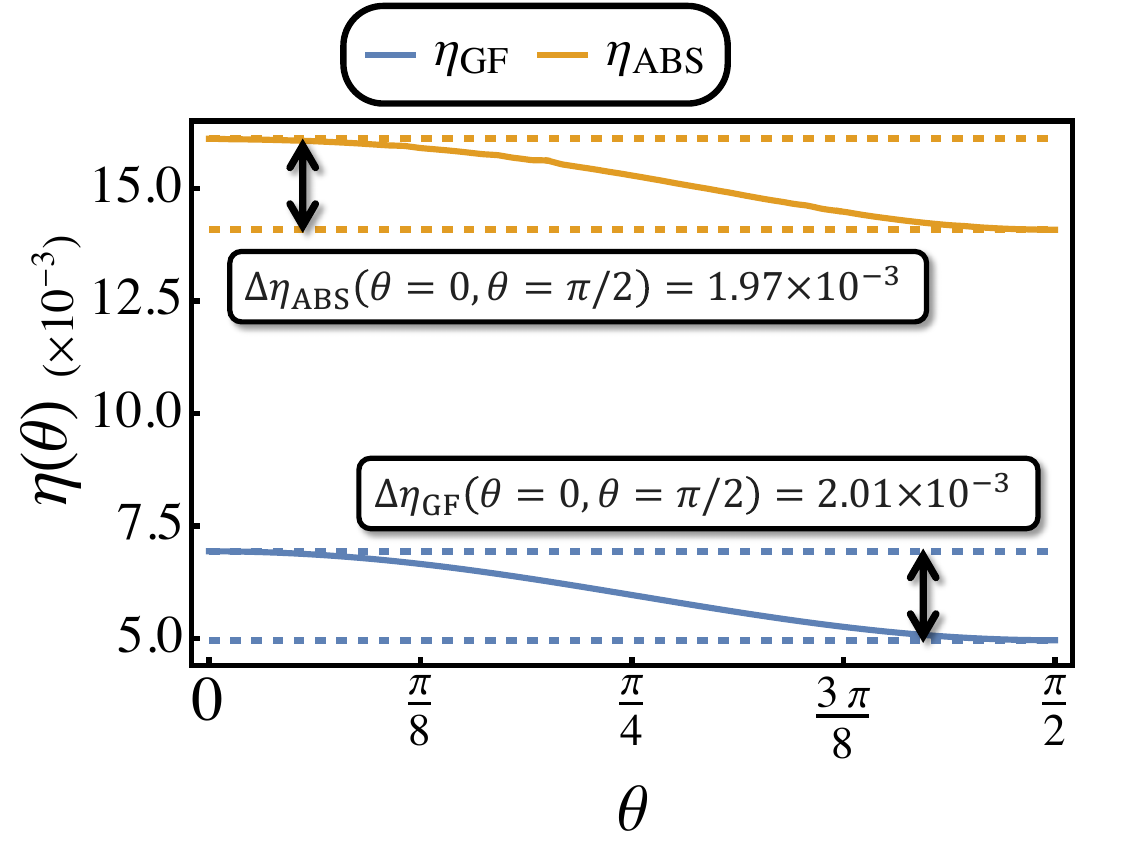}
		\caption{Comparison between the supercurrent computed with GF techniques and the supercurrent computed from the complex Andreev levels as in Eq.~\eqref{Jpol_T0_simp}. The system parameters are the same as in Fig.~\ref{fig: 8_NH_WC_ABS_F_lead_B_non_collinear} except for $\theta$.}
		\label{fig: 11_JInt_measure_vs_theta}
	\end{figure}
    
    In Fig.~\ref{fig: 11_JInt_measure_vs_theta}, we show the current norm for $J_{\rm ABS}$ and for $J_{\rm tot}$, that we here address as $\eta_{\rm ABS}$ and $\eta_{\rm GF}$ respectively.
    This norm quantifies the phase-averaged current flowing through the junction and it is chosen due to its stability in the presence of EPs. 
    
    By comparing $\eta_{\rm ABS}$ and $\eta_{\rm GF}$ we can evaluate the predictive power of the Andreev NH model. We observe that the overall angle dependence of the Josephson current is effectively captured by the quasi-ABS description.

	%

\end{document}